\documentclass[10pt,english,aps,prl,floats,manuscript,superscriptaddress,floatfix,twocolumn,preprint]{revtex4}
\usepackage[T1]{fontenc}
\usepackage[latin9]{inputenc}
\usepackage{bigints}
\usepackage[usenames,dvipsnames]{color}
\usepackage{float}
\usepackage{amssymb,amsmath}
\usepackage{graphicx}
\usepackage{epsfig}
\usepackage[normalem]{ulem}
\usepackage{booktabs}
\usepackage{setspace}
\usepackage{esint}
\usepackage{natbib}
\usepackage{bigints}
\usepackage{color}
\usepackage{babel}
\usepackage{amsmath,amsfonts,amssymb}
\usepackage{pifont}
\usepackage{rotating}
\usepackage{graphicx}    
\usepackage{epstopdf}
\usepackage{color}
\usepackage[caption=false]{subfig}
\usepackage{soul}
\usepackage{color}
\usepackage{babel}
\usepackage{array}
\usepackage{float}
\usepackage{booktabs}
\usepackage{setspace}
\usepackage{esint}
\usepackage[unicode=true,pdfusetitle,
bookmarks=true,bookmarksnumbered=true,bookmarksopen=false,
breaklinks=true,pdfborder={0 0 0},pdfborderstyle={},backref=false,colorlinks=true]
{hyperref}
\hypersetup{
	linkcolor=red, citecolor=blue,  urlcolor=blue}
\definecolor{dark-blue}{rgb}{0.15,0.15,0.4}

\makeatletter
\@ifundefined{textcolor}{}
{%
 \definecolor{BLACK}{gray}{0}
 \definecolor{WHITE}{gray}{1}
 \definecolor{RED}{rgb}{1,0,0}
 \definecolor{GREEN}{rgb}{0,1,0}
 \definecolor{BLUE}{rgb}{0,0,1}
 \definecolor{CYAN}{cmyk}{1,0,0,0}
 \definecolor{MAGENTA}{cmyk}{0,1,0,0}
 \definecolor{YELLOW}{cmyk}{0,0,1,0}
}

\newcommand{\SPEDITOKAY}[2]{{}{\textcolor{black}{#2}}}

\newcommand{\SPHIDE}[1]{{}}

\makeatother

\usepackage{babel}
\begin{document}
\title{Gapless quantum spin liquid in the triangular system
Sr$_{3}$CuSb$_{2}$O$_{9}$ }
\author{S. Kundu}
\email{skundu37@gmail.com}

\affiliation{Department of Physics, Indian Institute of Technology Bombay, Powai,
Mumbai 400076, India}
\author{Aga Shahee }
\affiliation{Department of Physics, Indian Institute of Technology Bombay, Powai,
Mumbai 400076, India}
\author{Atasi Chakraborty}
\affiliation{School of Physical Sciences, Indian Association for the Cultivation
	of Science, Jadavpur, Kolkata 700032, India}
\author{K. M. Ranjith }
\affiliation{Max Planck Institute for Chemical Physics of Solids, 01187 Dresden,
Germany}
\author{B. Koo}
\affiliation{Max Planck Institute for Chemical Physics of Solids, 01187 Dresden,
Germany}
\author{Jörg Sichelschmidt }
\affiliation{Max Planck Institute for Chemical Physics of Solids, 01187 Dresden,
Germany}
\author{Mark T.F. Telling}
\affiliation{ISIS Pulsed Neutron and Muon Source, STFC Rutherford Appleton Laboratory,
	Harwell Campus, Didcot, Oxfordshire OX110QX, United Kingdom}
\author{P. K.  Biswas}
\affiliation{ISIS Pulsed Neutron and Muon Source, STFC Rutherford Appleton Laboratory,
	Harwell Campus, Didcot, Oxfordshire OX110QX, United Kingdom}
\author{M. Baenitz}
\affiliation{Max Planck Institute for Chemical Physics of Solids, 01187 Dresden,
Germany}
\author{I. Dasgupta}
\affiliation{School of Physical Sciences, Indian Association for the Cultivation
of Science, Jadavpur, Kolkata 700032, India}
\author{Sumiran Pujari}
\affiliation{Department of Physics, Indian Institute of Technology Bombay, Powai,
Mumbai 400076, India}
\author{A. V. Mahajan}
\email{mahajan@phy.iitb.ac.in}
\affiliation{Department of Physics, Indian Institute of Technology Bombay, Powai,
Mumbai 400076, India}
\date{\today}
\begin{abstract}
We report gapless quantum spin liquid behavior in the layered triangular
Sr$_{3}$CuSb$_{2}$O$_{9}$ (SCSO) system. X-ray diffraction shows superlattice reflections associated
with atomic site ordering into triangular  Cu planes well-separated by Sb planes.  Muon
spin relaxation ($\mu$SR) measurements show that the $S = \frac{1}{2}$ moments at the magnetically active Cu  sites remain dynamic down to 65\,mK in spite of a large antiferromagnetic exchange scale evidenced by a large
Curie-Weiss temperature $\theta_{\mathrm{cw}} \simeq 
  -143$ K as extracted from the
bulk susceptibility. Specific heat measurements also show no sign of long-range
order down to 0.35 K. The magnetic specific heat  ($\mathit{C}$$_{\mathrm{m}}$)
below 5\,K reveals a $\mathit{C}$$_{\mathrm{m}}$ $=$ $\gamma T$ + $\alpha T$$^{2}$
behavior. The significant $T$$^{2}$ contribution to the magnetic specific heat invites a phenomenology in terms of the so-called Dirac spinon excitations with a linear dispersion. From the low-$T$ specific heat data, we estimate the dominant exchange scale to be  $\sim 36$ K using a Dirac spin liquid ansatz which is not far from the values inferred from microscopic density functional theory
calculations ($\sim 45$ K) as well as high-temperature susceptibility analysis ($\sim 70$ K). The linear specific heat coefficient is about 18 mJ/mol-K$^2$ which is somewhat larger than for typical Fermi liquids. 
\end{abstract}
\maketitle
\textit{Introduction}: The search for novel spin liquids has been
driving the community of quantum magnetism ever since the proposal
of Fazekas and Anderson \cite{Fazekas1974}.  It is now theorized that spin liquids come in various flavors,
may have gapped or gapless excitations, may be topological or not \cite{Zhou2017}.
Frustration disfavors magnetic order and is thus generically sought
as a route to realizing spin liquids. Geometric frustration from the
lattice composed of triangular motifs forms a key class
in this search going back to the Fazekas-Anderson work for the triangular
lattice \cite{Fazekas1974}. While the $S = \frac{1}{2}$ uniform triangular lattice
 (with nearest-neighbor Heisenberg exchange $J_{1}$) has  120$^{\circ}$
\SPEDITOKAY{}{long-range} order \cite{Kadowaki1995,Ishii2011,Shirata2012} also supported by theory 
\cite{Huse1988,Jolicoeur1989,Singh1992,Bernu1994,Capriotti1999}, the  presence of \SPEDITOKAY{a next-nearest neighbor 
$J_{2}$}{further neighbor exchanges} \SPEDITOKAY{provides a region in the phase diagram ($0.06\,\lesssim\,J_{2}/J_{1}\,\lesssim\,0.17$)
where a quantum paramagnet is predicted to stabilize}{has been argued to enhance
frustration and induce spin liquid behavior} \cite{Kaneko2014, Zhu2015,Iqbal2016}.  
\SPEDITOKAY{This}{One candidate} state for such a spin liquid has linearly dispersing low-energy excitations \cite{Knolle2019} and has been dubbed as a Dirac quantum spin liquid (QSL).
They lead to a $T^{2}$-variation of the heat capacity in zero magnetic field, and an additional linear-in-$T$ variation 
in the  presence of a magnetic field. For $S = \frac{1}{2}$ triangular
lattices, previously reported spin liquids \cite{Shimizu2003,Itou2007,Itou2008,Yamashita2008,Paddison2017,Baenitz2018, Sarkar2019}
have not shown a $T^{2}$ behavior in the specific heat. Here we report on the finding
of such a candidate Dirac QSL in the Sr$_{3}$CuSb$_{2}$O$_{9}$ (SCSO) triangular lattice system.

\begin{figure}[h]
	\centering{}\includegraphics[scale=0.28]{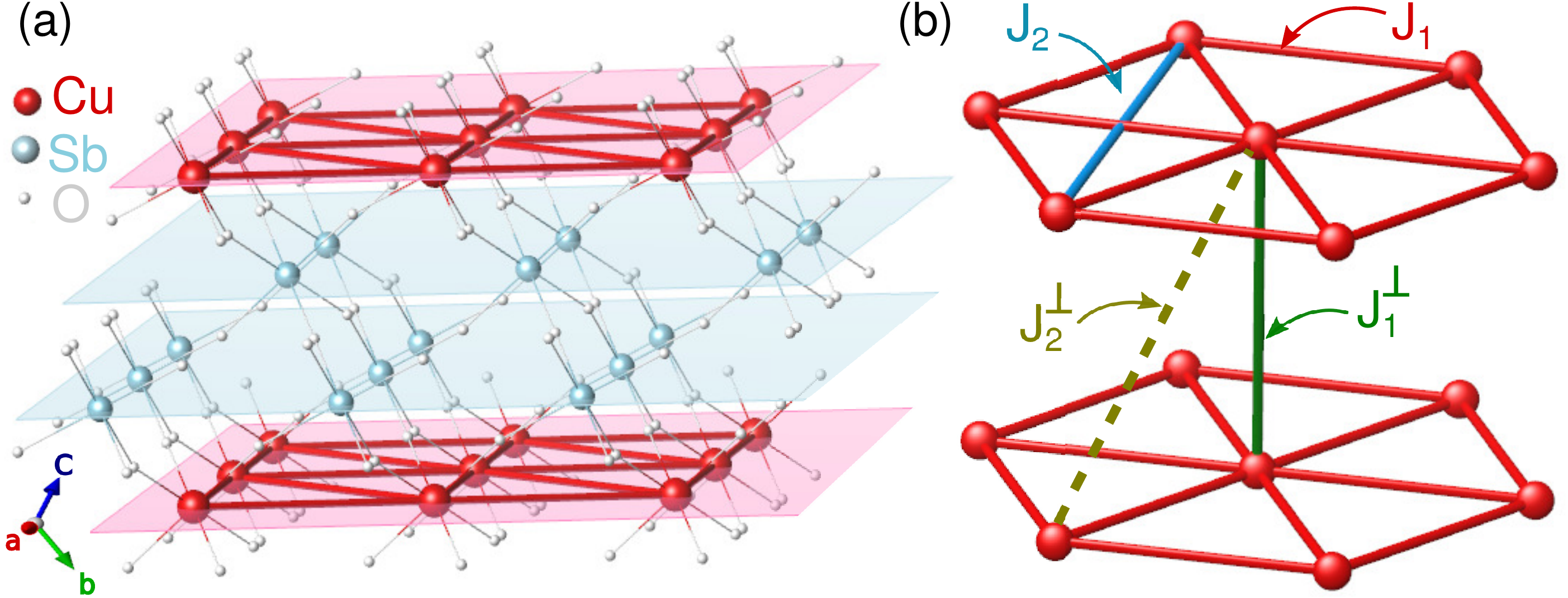}\caption{\label{fig:scso-schematic}{\small{}(a) A schematic of the SCSO crystal structure is shown highlighting the individual planes of Cu (red) and Sb (blue) atoms. The Sr atoms are omitted for visual clarity. (b) The paths corresponding to different exchange couplings.}}
\end{figure}

Taking inspiration from the  triple perovskite Ba$_{3}$CuSb$_{2}$O$_{9}$ (of the form A$_3$B$_3$X$_9$) with a hexagonal lattice  which 
has been inferred to be 
\SPEDITOKAY{an}{a}  $S = \frac{1}{2}$ QSL \cite{Zhou2011,Ishiguro2013}, we considered
the possibility of replacing the Ba$^{2+}$ by the smaller Sr$^{2+}$
ion resulting in chemical pressure and concomitant effects on the 
crystal structure and the magnetic ground state. Ref. \cite{Bush2008} has
reported on the dielectric properties of 
Sr$_{3}$CuNb$_{2}$O$_{9}$ while SCSO is largely
unexplored. These triple perovskites crystallize in the tetragonal crystal system 
which is different from that of Ba$_{3}$CuSb$_{2}$O$_{9}$. 
Significantly different ionic sizes and charges  of Sb$^{5+}$ (0.60 \AA) and Nb$^{5+}$ (0.64 \AA) compared to  Cu$^{2+}$ (0.73 \AA)  
should favor atomic site ordering at the B-sites. 
This has also been seen in homologous compounds Sr$_3$CaIr$_2$O$_9$ \cite{Wallace2015} and Sr$_3$CaRu$_2$O$_9$ \cite{Rijssenbeek2002}).
With 1:2 ordering at the B-site, the (111) planes (pertinent to the pseudo-cubic
lattice) will have successive Cu planes with an edge-shared triangular
geometry separated by two Sb/Nb planes (see Fig. \ref{fig:scso-schematic} for a schematic). X-ray diffraction indeed shows superlattice peaks supporting this site ordering in SCSO. 

Our experiments on SCSO have shown the following salient results:
1) The bulk susceptibility of SCSO shows
no sign of long-range order (LRO) down to 1.8 K. It shows Curie-Weiss behavior
with a Curie-Weiss temperature $\theta_{\mathrm{cw}} \simeq$
-143\,K. There is no bifurcation in
the zero-field cooled (ZFC) and field cooled (FC) magnetization in low
field either. 
2) Zero-field muon spin relaxation
(ZF-$\mu$SR) data reconfirm the absence of any magnetic ordering down to 65 mK. Longitudinal-field $\mu$SR (LF-$\mu$SR) data show that the moments remain dynamic down to very low temperatures.
3) Magnetic heat capacity data are also devoid of any indications
of a phase transition down to 0.35 K. A $\gamma T + \alpha T^2$ form is 
seen in the low-$T$ behavior. 

\SPEDITOKAY{The}{An} unequivocal presence of a $T^2$ contribution in the magnetic
	specific heat, in the absence of any evidence of magnetic order,  
naturally suggests a phenomenology in terms of a Dirac QSL. Density functional theory
(DFT) calculations for the B-site ordered structure also support a model of SCSO as a \SPEDITOKAY{``$J_{1}$-$J_{2}$"}{} 
triangular lattice quantum antiferromagnet \SPEDITOKAY{}{with small further-neighbor frustrating 
antiferromagnetic exchanges}
(in the  (111) planes pertinent to the pseudo-cubic cell) for which a Dirac QSL has been 
predicted \cite{Kaneko2014,Zhu2015,Iqbal2016}.  

\begin{figure}[h]
	\centering{}\includegraphics[scale=0.35]{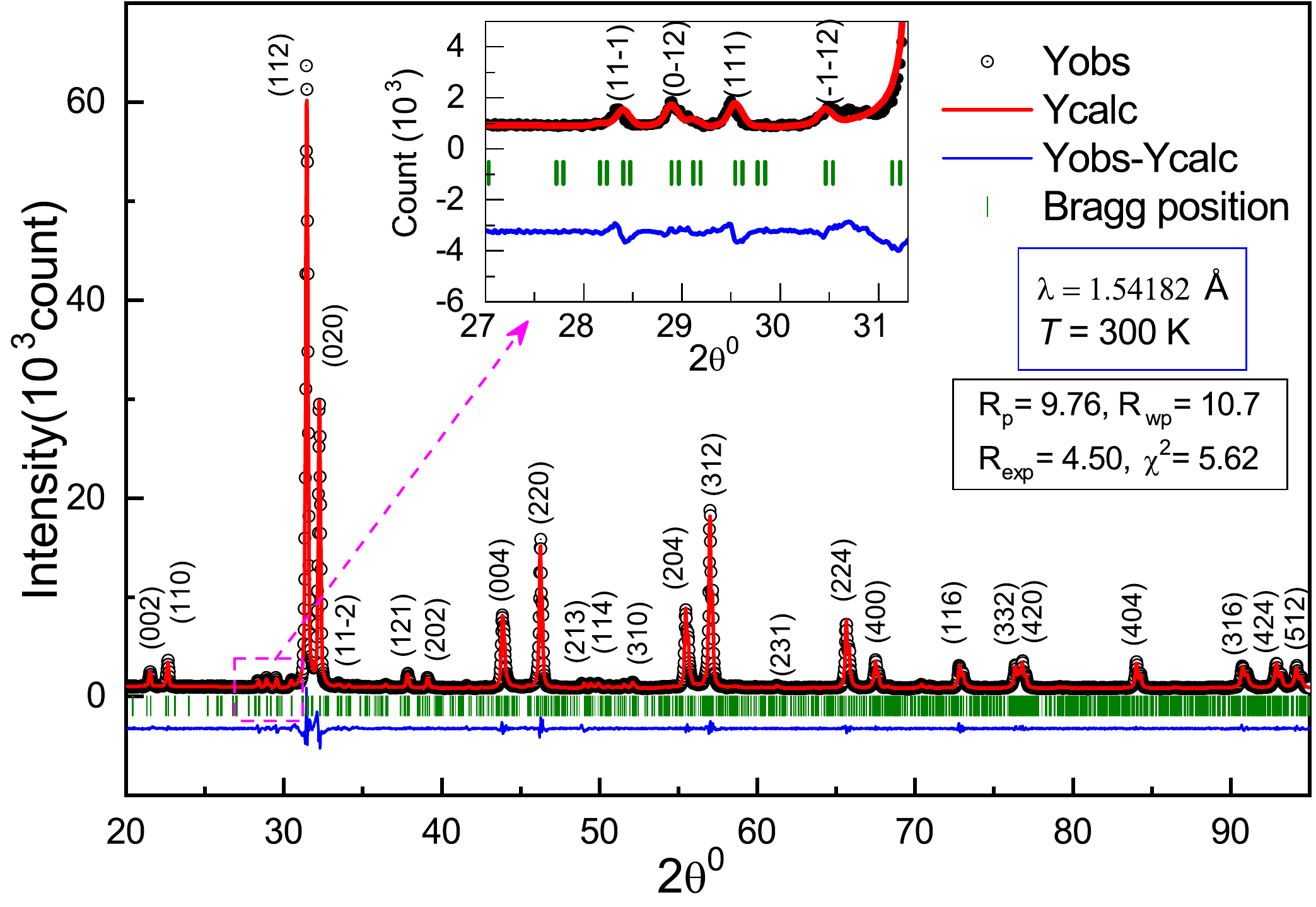}\caption{\label{fig:Reitveld_superlattice 1/3rd}{\small{}Rietveld refinement	of Sr$_{3}$CuSb$_{2}$O$_{9}$ is shown along with Bragg positions
			and corresponding Miller indices (hkl) with K-modulation ($\frac{1}{3}$,
			$\frac{1}{3}$, $\frac{1}{3}$) (referred to a pseudo-cubic cell) to fit the superlattice peaks.}}
\end{figure}

\textit{Results and discussion}: 
The Rietveld refinement of XRD data on our polycrystalline SCSO 
sample by Fullprof suite software \cite{Carvajal1990} is shown in
Fig. \ref{fig:Reitveld_superlattice 1/3rd}. Fitting with a body-centered tetragonal structure with space group
$I{\rm 4}/mcm$ (140) (shown in SM \cite{SM-SCSO}) does not account for a few peaks at low
angles (region shown in the inset of Fig. \ref{fig:Reitveld_superlattice 1/3rd}).
1:2 atomic site ordering refines the data well with a propagation vector $\vec{K}$
= ($\frac{1}{3}$, $\frac{1}{3}$, $\frac{1}{3}$). The obtained lattice parameters
are $\mathit{a}$ = $\mathit{b}$ = 5.547 \AA,  $\mathit{c}$
= 8.248 \AA.   

\begin{figure}[h]
\centering{}\includegraphics[scale=0.30]{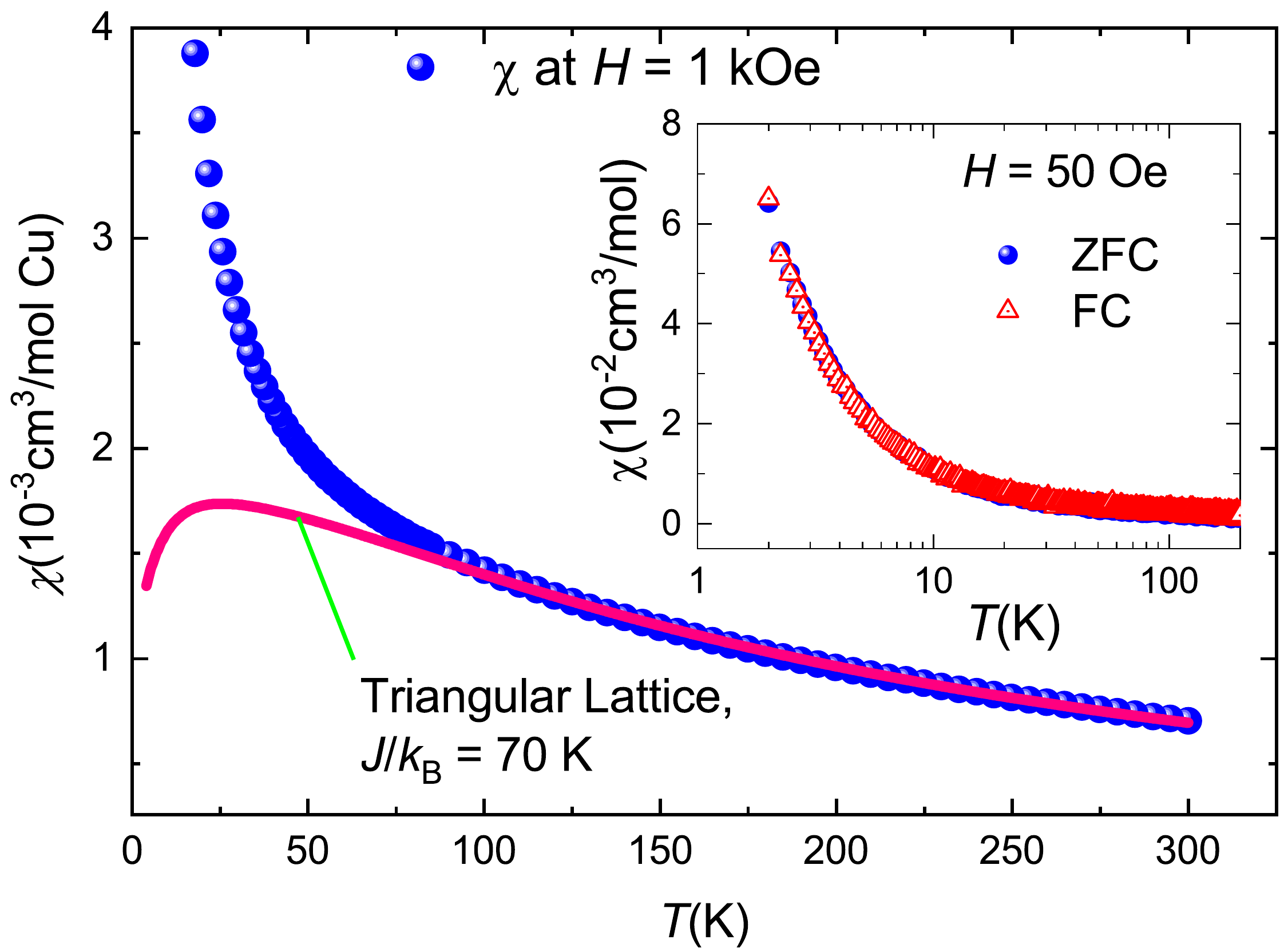}\caption{\label{fig:Curie-Weiss-fit-of SCSbO}{\small{}The variation of $\mathit{\chi}(T)$ for SCSO in $H$ = 1 kOe is shown in the main figure. The solid line is a theoretical TLAF fit for  data in the range 150-300 K and then extrapolated down to 2 K. In the inset,
no bifurcation is seen in the plot of $\chi(T)$ vs $\mathit{T}$
between ZFC and FC data in
 $\mathit{H}$ = 50 Oe.}}
\end{figure}

Fig. \ref{fig:Curie-Weiss-fit-of SCSbO} shows the DC susceptibility,
$\chi (T)$ ($=$$\frac{M(T)}{H}$) of SCSO in $\mathit{H}$ =
1\,kOe. No indications of long-range  order are seen down to 2\,K. We fit the data
(in the \SPEDITOKAY{}{high-temperature} range of $150-300$\,K) to the theoretical Triangular Lattice  Antiferromagnet (TLAF) model 
for a 2D spin-1/2 system \cite{Elstner_Singh_Young1993} using  $\chi=\chi_{0}+ \chi_{\rm {TLAF}}$
as in Ref. \cite{Tamura_Kato2002,Haraguchi_etal2015}.  Here $\chi_{0}$
is the temperature independent susceptibility (arising from core diamgnetism and a van Vleck contribution), $C=N_{A}g^{2}\mu_{B}^{2}/4k_{B}$ is the Curie constant, and $J_1$ is the nearest neighbour exchange coupling \cite{corrections}. 
Fixing $\mathit{C}$ = 0.375\,K cm$^{3}$/mol Cu for our $S = \frac{1}{2}$ system, 
this TLAF fit yields $\chi_{0}$= -2.10$\times$ 10$^{-4}$ cm$^{3}$/mol Cu, 
and  $|J_1/k_{B}|$ = 70\,K \cite{cwfootnote}. 
   
The inset
of Fig. \ref{fig:Curie-Weiss-fit-of SCSbO} shows the ZFC/FC susceptibility.
No bifurcation is seen down to 2 K. AC susceptibility data (see
SM \cite{SM-SCSO}) show no anomaly either. This rules out the existence of spin-glass
behavior in the system \cite{esrfootnote}. 

\begin{figure}[h]
\centering{}\includegraphics[width=0.85\linewidth,trim=20mm 10mm 0mm 10mm,clip=true]{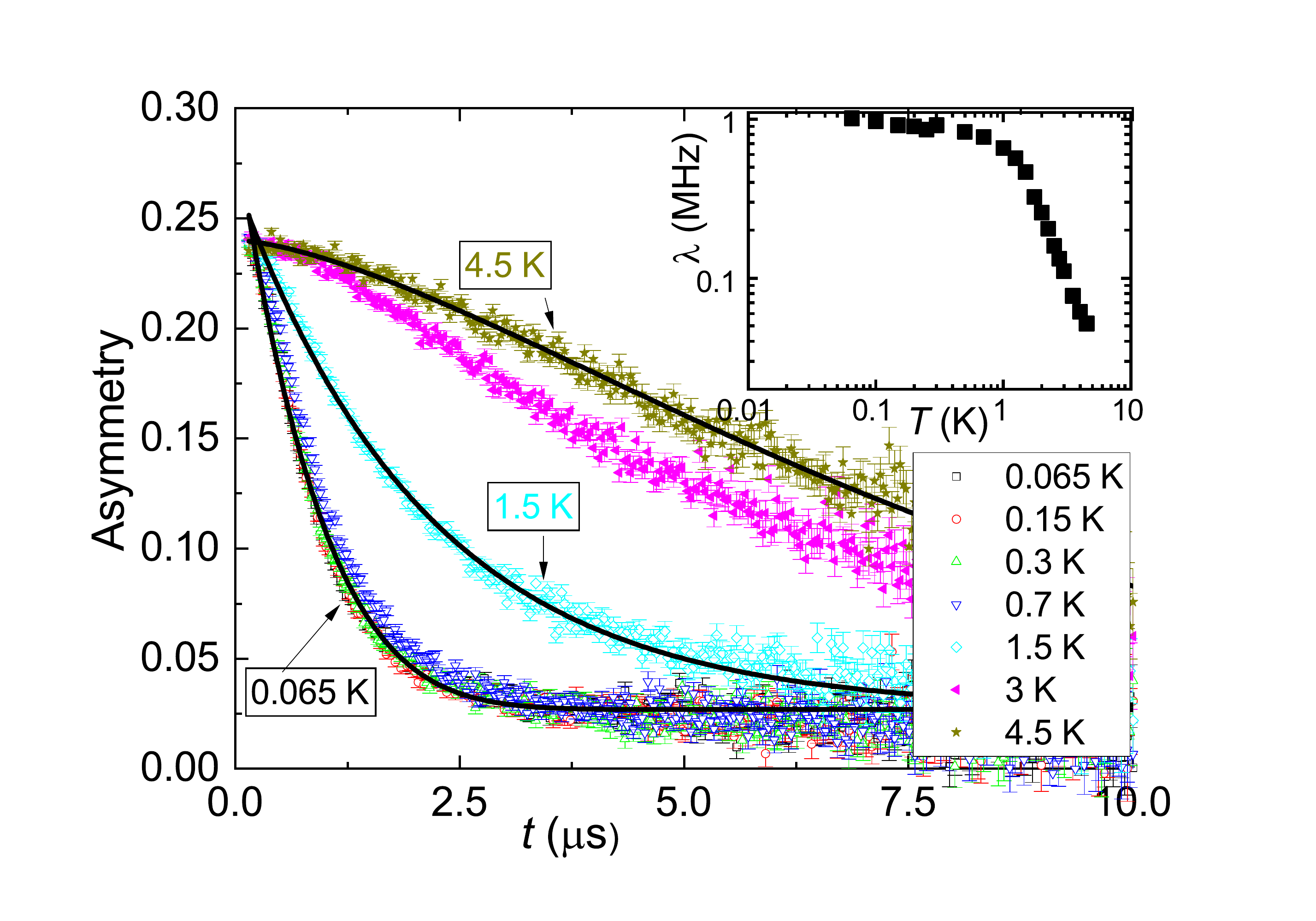}
\caption{\label{fig:30}The variation of the muon asymmetry with time is shown
at various temperatures in zero field for Sr$_{3}$CuSb$_{2}$O$_{9}$. On fitting these data as described in the text, the muon relaxation rate was obtained and is shown in the inset. }
\end{figure}

Additional evidence for the absence of static magnetism down to even lower
temperatures (65 mK) was gathered from $\mu$SR
experiments. The ZF muon asymmetry was measured as a function of $T$
between 65 mK and 4 K. Strong relaxation of the muon is seen at the
lowest temperature and we fit the time dependence of the asymmetry
to $\mathit{A\mathrm{(}t\mathrm{)}}=A_{0}G_{KT}(\Delta,t)e^{-\lambda(T)t}+A_{bg}$.
Here, $\mathit{G_{KT}\mathrm{(}\Delta,t\mathrm{)}}$ is the static
Kubo-Toyabe Gaussian function coming from relaxation due to the
nuclear moments whereas the exponential decrease is from the relaxation
due to electron moments. $\mathit{A_{bg}}$ is the constant background signal (due to a small fraction of the muons missing the sample and hitting the sample holder and cryostat wall)
and $\mathit{A_{\mathrm{0}}}$ is the initial muon asymmetry. The
muon asymmetry data at various $T$ are shown in Fig. \ref{fig:30}.
The absence of oscillations in the muon asymmetry data suggests the absence of  any short/long-range magnetic ordering down to 65 mK. Further, the absence of the 1/3rd tail suggests the
dynamic nature of the electronic spins.  The LF-$\mu$SR data will further validate the dynamic nature of the moments.  

As seen in Fig. \ref{fig:30}, the relaxation curves are essentially
unchanged from 65 mK to about 1 K and at higher temperatures, the
muon depolarisation rate decreases. From the above analysis, the muon
relaxation rate due to electron moments is obtained as shown in the inset of Fig.
\ref{fig:30}. It is seen that there is a gradually faster relaxation
of muons with a lowering of temperature. However, there is no large
increase or a critical divergence and  the electron
moments remain dynamic till the lowest temperatures.

We have also monitored muon relaxation in longitudinal fields and
find that even in our highest field of 3200 Oe, residual relaxation
is still present (see SM \cite{SM-SCSO}). The field dependence is remarkably similar to that of YbMgGaO$_4$ \cite{Li2016} with a local moment fluctuation frequency of about
18 MHz and the presence of long-time spin correlations. The qualitative and
quantitative outcomes of $\mu$SR  on SCSO are typical
of other QSL.

\begin{figure}[h]
	\begin{centering}
		\includegraphics[width=0.85\linewidth]{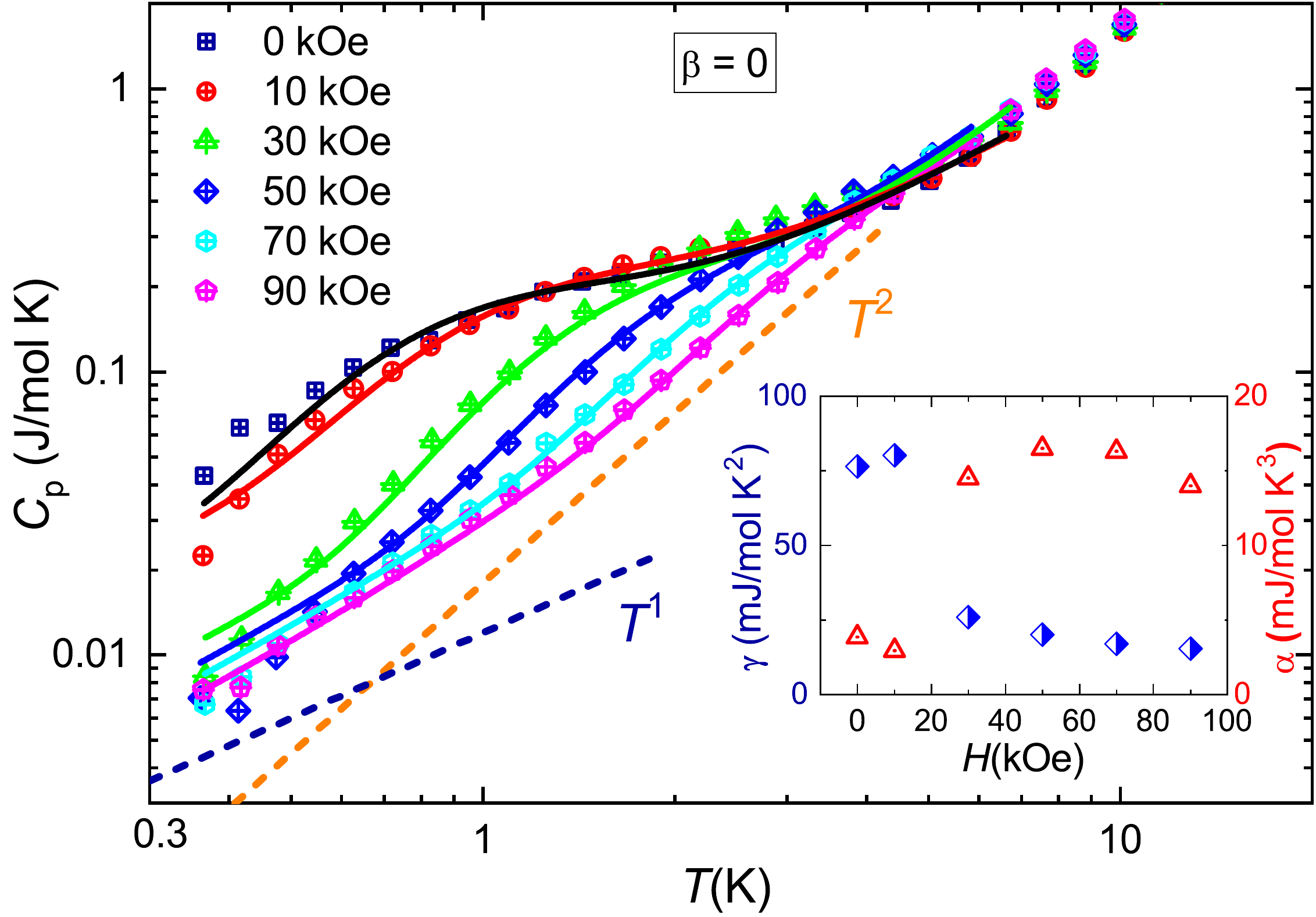}
		\par\end{centering}
	\caption{\label{fig:heat capacity}{The heat capacity 
			of SCSO is plotted as a function of temperature in various fields. The solid lines are fits as explained in the text. Dashed lines represent power law variations $T^1$ and $T^2$. Inset:
			The variation of the coefficients of the linear (blue diamonds; left $y$-axis) and quadratic (red triangles; right $y$-axis) terms is shown as a function of $H$. }}
\end{figure}

To further rule out LRO and to probe the nature of low-energy excitations,
we measured the heat capacity of the sample $C_{\mathrm{p}}(T)$ in
different fields (0-90\,kOe) in the $T$-range (0.35 - 200)\,K.  
As shown in Fig. \ref{fig:heat capacity}, a hump is seen in the heat capacity 
which moves to higher temperatures with increasing field, and can be ascribed 
to the Schottky anomaly commonly seen in many quantum magnets \cite{Vries2008,Zhou2011,Han2014}. 
This arises from a small fraction of free spins 
(either of extrinsic origin or from edge spins of correlated regions). In the low-$T$
regime, the scaling is observed to be close to $T^2$. This already hints at the presence 
of linearly-dispersing excitations. 
We thus model the 
specific heat below 6 K as 
$C_{\mathrm{p}}(T,H) = \gamma(H) T + \alpha(H) T^2 + fC_\mathrm{Schottky}(T,H)$ 
motivated by a Dirac QSL phenomenology. 
$C_\mathrm{Schottky}$ is the standard Schottky contribution of a two-level system (see SM \cite{SM-SCSO}) 
with $f$ being the fraction of free $S = \frac{1}{2}$ entities.  
Solid lines in Fig. \ref{fig:heat capacity} show fits to the as-measured 
$C_{\mathrm{p}}$  with the parameter values  $\gamma_{avg} = 18$ mJ/mol-K$^2$, $\alpha_{avg} = 15$ mJ/mol-K$^3$, and  $f = 2.5$\%. For $\gamma$ and $\alpha$,
the quoted average is over the fits for $H \geq 30$ kOe. 
The Schottky fraction $f$ was obtained from the fit for 90 kOe data and 
kept fixed for other fields. A $\beta T^3$ contribution from the lattice can also be included in the analysis, 
but it makes no essential difference to the fits. 
Fits at high-$T$ to the lattice heat capacity help us fix 
$\beta = 3 \times 10^{-4}$ J/mol-K$^4$ (see SM \cite{SM-SCSO}). 
As the lattice contribution in the low-$T$ range is negligible, 
we get very similar parameter values $\gamma_{avg} = 16$ mJ/mol-K$^2$, $\alpha_{avg}  = 15$ mJ/mol-K$^3$ with $f = 2.5\%$. 
Most importantly, 
the fitting significantly worsens \emph{without} a $T^2$ component and can not be considered as fitting the data (see SM \cite{SM-SCSO}).

As far as  $S = \frac{1}{2}$ triangular lattice spin liquids are concerned, a linear variation of the low-temperature
specific heat was observed in $\kappa$-(BEDT-TTF)$_2$Cu$_2$(CN)$_3$ \cite{Yamashita2008} and Ba$_3$CuSb$_2$O$_9$ \cite{Zhou2011}.  
This was interpreted in terms of a spinon Fermi surface. However, the $T^{2}$ term in $C_{\mathrm{m}}(T)$ observed at low-$T$ in SCSO is perhaps
the first such observation in triangular $S=\frac{1}{2}$ magnets. From this crucial presence of $T^2$ contributions, 
we infer the presence of gapless excitations
with a linear dispersion \cite{T2HCfootnote}. Within a Dirac QSL phenomenology that naturally gives such
excitations, we can 
estimate the magnetic exchange strength from low-temperature specific heat data to
compare with previous estimates from high-temperature data.  We take the following effective mean-field 
Hamiltonian as our Dirac QSL ansatz
$\mathcal{H}  = \frac{J_{\text{eff}}}{2} \sum_{\langle i,j \rangle \in \triangle \text{ lattice}, \sigma}
\chi_{ij} c^\dagger_{i,\sigma} c_{j,\sigma} + \text{ h.c.}$
such that $\pi$-flux is inserted in the up triangles \cite{Iqbal2016, Lu2016}. $J_{\text{eff}}$ is expected to be the same order as
the dominant exchange ($J_{\text{eff}} \sim \mathcal{O}(J_1)$)
Near the single Dirac cone in the Brillouin zone (see SM \cite{SM-SCSO} for details),
the effective low energy spectrum is
$\epsilon_{\mathbf{k}} = \pm \sqrt{\frac{3}{2}} J_{\text{eff}} \: k$. Therefore at low temperatures
($T \rightarrow 0$), we obtain 
$\frac{C(T)}{k_B}  = 2.292 \frac{(k_B T)^2}{J_{\text{eff}}^2}$.
In presence of Zeeman coupling to an external magnetic field, this
gets modified to $\frac{C(T)}{k_B}  \approx 2.292 \frac{(k_B T)^2}{J_{\text{eff}}^2} + 0.696 \frac{(k_B T) \left(\frac{g \mu_B |H|}{2}\right)}{J_{\text{eff}}^2}$
as $\frac{g \mu_B |H|}{2 k_B T} \gg 1$,
which is indeed the functional form used earlier for fits.
From the coefficient $\alpha_{avg}$ of the $T^2$ contribution, we estimate $\frac{J_1}{k_B} \sim 36$ K. This completely independent low-$T$, macroscopic or thermodynamic
estimate for $J_1$ is of the same order as the previous estimate from high-$T$ susceptibility data
which we consider significant \cite{dsl_pheno_details2}.
In fact, from the low field values of $\alpha$, the estimate is $\frac{J_1}{k_B} \sim 76$ K.
Furthermore, these estimates also agree with microscopic estimates based on
a DFT study of the SCSO perovskite system to be discussed soon. 

We finally note an anomaly that is present in the heat capacity data at low fields,
seen as a ``separation'' between the curves as we go from 10 kOe to 30 kOe in Fig. \ref{fig:heat capacity}.
This is also reflected in the fits based on our linearly-dispersing Dirac QSL ansatz as a jump in $\alpha$ and $\gamma$ 
at these fields as seen in the inset of Fig. \ref{fig:heat capacity}. 
Note that the Schottky gap is non-zero even in the absence of a magnetic field for 
SCSO (see SM \cite{SM-SCSO}) which is a feature seen in other QSLs as 
well \cite{Vries2008,Zhou2011,Han2014} and is ascribed to interaction of the 
orphan spins with the correlated regions. 
We speculate this happens when the orphan spins suddenly decouple from the 
correlated regions as the applied field exceeds $\sim$ 10 kOe (which might 
be the effective internal field strength of their interaction). 
Since this ``separation" in
the data sets at low field is observed in the raw or as-measured $C_p$ data
prior to any fitting, 
we think that it has a distinct physical origin
(leading to an artificial jump in $\alpha$, $\gamma$ parameters when using a fitting form
based on assuming only linear-dispersing excitations). 
There could be other explanations for this observation as well.
Also for $H \geq $ 30 kOe, a slight decrease of $\gamma$ with field is found from the data whereas an increase is predicted by our ansatz, while $\alpha$ remains roughly independent
of field as predicted by our ansatz. 
Nonetheless, another supporting feature in the data is that the ratio of the quadratic 
and linear components agrees with the theoretical value
quite well for $H$ = 50 kOe and to within a factor of 3 
for all $H \geq $ 30 kOe data sets despite the anomalous field dependence of $\gamma$.
The deviations  suggest that something 
else may also be contributing to the heat capacity in addition to the linearly dispersing
excitations. \SPEDITOKAY{}{We remark here that 1)
$\frac{g \mu_B |H|}{2 k_B T} \gg 1$ is not strictly true for the applied fields,
and 2)
that our ansatz is at a mean-field level, though the $T^2$ behavior is robust
to beyond mean-field effects for a Dirac QSL. \cite{Ran_etal2007,Kim_Lee_Wen1997}}

To check the viability of a Dirac QSL phenomenology argued above
\SPEDITOKAY{based on}{as arising from} a 
\SPEDITOKAY{$J_1$-$J_2$}{} triangular quantum magnet \SPEDITOKAY{}{with further-neighbor frustrating
antiferromagnetic exchanges} \cite{Kaneko2014,Iqbal2016,Zhu2015}, 
we have carried out first principles electronic structure calculations 
based on DFT for the experimentally determined triple perovskite structure of 
Sr$_3$CuSb$_2$O$_9$.
%
All the electronic structure calculations are carried out using 
DFT in the pseudo-potential plane-wave basis within the  generalized gradient approximation
(GGA) \cite{Perdew_G_1996} supplemented with Hubbard U \cite{Anisimov_U}
as encoded in the Vienna \textit{ ab~initio} simulation package (VASP) \cite{Kresse_A_1993,Kresse_E_1996}
with projector augmented  wave potentials \cite{Blochl_P_1994,Kresse_F_1999}.  The calculations are done
with standard values of $U_{\mathrm{eff}} \equiv U_{\mathrm{d}}-J_{\mathrm{H}} = 6.5$ eV \cite{Anisimov_U}
chosen for Cu. 
The non-spin-polarized total and Cu-d partial
density of states (DOS) for SCSO (see SM \cite{SM-SCSO}) reveal that the Fermi level is dominated by partially filled Cu e$_{g}$ states. The oxygen $O-p$ states are completely occupied while Sb $s$ and $p$ states and Sr $s$ states are completely empty consistent with the nominal ionic formula, Sr$^{2+}_{3}$Cu$^{2+}$Sb$^{5+}_{2}$O$^{2+}_{9}$ of this compound.
Spin polarized \SPEDITOKAY{GGA+U}{} 
calculations with ferromagnetic (FM) arrangement of  Cu spins yields total moment 1.0 $\mu_{B}$ per formula unit,
which further supports the $S = \frac{1}{2}$ moment of Cu and is consistent with the experimentally 
determined effective moment
($\mu_{\mathrm{eff}}= 1.72 \mu_{B}$). However, the calculated  magnetic moment per Cu site is 0.80 $\mu_{B}$, 
while 
rest of the moment is hosted on the ligand sites \SPEDITOKAY{(0.02, 0.04 $\mu_{B}$/ O)}{} 
due to substantial hybridization of Cu with ligands. \\

%
To estimate the exchange interactions, we consider the following Hamiltonian
$H=
J_{1} \sum^{\text{intra}}_{\langle i,j\rangle} \mathbf{S}_{i}. \mathbf{S}_{j}
+J_{2}\sum^{\text{intra}}_{\langle\langle i,j\rangle\rangle} \mathbf{S}_{i} . \mathbf{S}_{j}
+J_{1}^{\perp}\sum^{\text{inter}}_{\langle i,j\rangle} \mathbf{S}_{i}. \mathbf{S}_{j}
+J_{2}^{\perp}\sum^{\text{inter}}_{\langle i,j\rangle} \mathbf{S}_{i}. \mathbf{S}_{j}$.
\label{tothamEqn} 
Here $J_1$, $J_{2}$, $J^{\perp}_{1}$ and $J^{\perp}_{2}$ are respectively the nearest-neighbor (nn), 
the next-nearest-neighbor (nnn) intra-layer and (two distinct) 
inter-layer Heisenberg exchange parameters.  
We have calculated them 
employing the  ``Four State'' method \cite{Whangbo_P_2011, Kumar_S_2019}. 
This is based on a computation of the total energy of the
system with collinear spin alignment, where the spin configuration on two chosen sites \SPEDITOKAY{within i
a given unit cell} are modified while restricting rest of the spins to a base configuration.
Our calculations reveal that the estimates of the exchange interactions 
change up to 15$\%$ depending on the chosen base configuration. 
We find $J_1 \sim 3.92$ meV ($J_1/k_B \sim  45$ K) to be the dominant antiferromagnetic
exchange in this system.
This microscopic estimate agrees well with the values extracted from the high-temperature
susceptibility data and low-temperature specific heat data.
$J^{\perp}_{1} \sim 0.21 $ meV and $J^{\perp}_{2} \sim$ 0.11 meV are found be sub-dominant and 
robustly antiferromagnetic, thus adding to the frustration.
$J_{2} \sim 0.05$ meV was even smaller, and its sign also depended on 
the chosen base configuration. This last estimate is at the edge
of the accuracy of our DFT calculations ($\sim 0.05$ meV). 
The strength of the further-neighbor exchanges
are relatively much smaller because the Cu atoms are far apart, 
as well as the intermediate Sb atoms are smaller in size. 
Our first principles DFT results corroborates well with the experimental results, 
and lends credence to the scenario of spin liquid behavior 
induced by further-neighbor frustrating exchanges \cite{Kaneko2014,Zhu2015,Iqbal2016}.

\textit{Summary}: Our comprehensive set of data on SCSO show a 1:2 site ordering between 
Cu and Sb ions (giving rise to nearly isolated $S = \frac{1}{2}$ triangular planes) and 
a lack of LRO together with the presence of dynamic moments down to the lowest 
temperatures (65 mK, which is well below $\theta_{\mathrm{cw}}$=
-143\,K). In contrast to other Kagome or pyrochlore based QSL, here
the magnetic heat capacity at low-$T$ follows $\mathit{C}$$_{\mathrm{m}}$ $=$ $\gamma T$ + $\alpha T$$^{2}$.
The $T$$^{2}$ behavior of $\mathit{C_{m}}$ evinces the presence of gapless excitations
with a linear dispersion, and the dominant exchange extracted through such a phenomenology agrees quantitatively
with those from high-temperature susceptibility data and DFT estimates.
We believe that the presence of further-neighbor antiferromagnetic exchanges
induces Dirac QSL behavior in the triangular lattice system
of SCSO. 
Our work offers new \SPEDITOKAY{}{material} directions to explore in the field of QSLs. 
Further work to unravel the magnitude of the further-neighbor couplings or possibly ring exchange terms \cite{Motrunich2005, Grover2010} in promoting the spin liquid state is clearly warranted.

\section{acknowledgment}

We thank MHRD and Department of Science and Technology, Govt. of India
for financial support. 
We also thank Nandini Trivedi and Subhro Bhattacharjee for useful discussion and late Christoph Klausnitzer for technical support. 
Experiments at the ISIS Neutron and Muon Source
were supported by a beam-time allocation RB1910129 from the Science
and Technology Facilities Council.
SP acknowledges financial
support from Industrial Research and Consultancy Centre, IIT Bombay (17IRCCSG011) and
SERB, DST, India (SRG/2019/001419). 
IDG acknowledges financial support from  Technical Research Centre, DST and SERB India.
AVM, IDG and SP acknowledge hospitality and support of ICTS and APCTP during the 2nd
Asia Pacific Workshop on Quantum Magnetism (Code:ICTS/apfm2018/11).
\bibliographystyle{apsrev4-1}
\bibliography{SCSO}

\begin{thebibliography}{57}
\bibitem [{cor()}]{corrections}%
  \BibitemOpen
  \href@noop {} {}\bibinfo {note} {Sub-dominant exchange couplings are assumed
  to not affect the susceptibility significantly in the \emph{high-temperature}
  regime which is primarily controlled by the dominant exchange
  coupling.}\BibitemShut {Stop}%
\bibitem [{cwf()}]{cwfootnote}%
  \BibitemOpen
  \href@noop {} {}\bibinfo {note} {A CW fit at high-$T$, yields $\chi_{0}$ =
  -1.61$\times$ 10$^{-4}$ cm$^{3}$/mol Cu, $\mathit{C}$ = 0.384\,K cm$^{3}$/mol
  Cu and the $\theta_{\rm CW}$ = -143\,K. The inferred effective moment
  $\mu_{\rm eff}$ = 1.75 $\mu_{\rm B}$ is close to the expected value of 1.73
  $\mu_{\rm B}$ for S = 1/2 Cu$^{2+}$ ion with $g = 2$. A negative $\theta_{\rm
  CW}$ indicates antiferromagnetic correlations between the Cu$^{2+}$
  ions.}\BibitemShut {Stop}%
\bibitem [{esr()}]{esrfootnote}%
  \BibitemOpen
  \href@noop {} {}\bibinfo {note} {Similar conclusions were drawn from our
  electron spin resonance (ESR) and nuclear magnetic resonance (NMR)
  experiments (see SM \cite{SM-SCSO}).}\BibitemShut {Stop}%
\end{thebibliography}
\begin{thebibliography}{8}
\bibitem [{che()}]{chem-pot}%
  \BibitemOpen
  \href@noop {} {}\bibinfo {note} {We make the usual observation that changes
  in the chemical potential due to finite temperatures causes at most
  subleading order corrections.}\BibitemShut {Stop}%
\end{thebibliography}

\end{document}


\title{Supplementary Material for Gapless quantum spin liquid in the triangular system
	Sr$_{3}$CuSb$_{2}$O$_{9}$}
\author{S. Kundu}
\email{skundu37@gmail.com}
\affiliation{Department of Physics, Indian Institute of Technology Bombay, Powai,
Mumbai 400076, India}
\author{Aga Shahee }
\affiliation{Department of Physics, Indian Institute of Technology Bombay, Powai,
Mumbai 400076, India}
\author{Atasi Chakraborty}
\affiliation{School of Physical Sciences, Indian Association for the Cultivation
of Science, Jadavpur, Kolkata 700032, India}
\author{K. M. Ranjith}
\affiliation{Max Planck Institute for Chemical Physics of Solids, 01187 Dresden,
Germany}
\author{B. Koo}
\affiliation{Max Planck Institute for Chemical Physics of Solids, 01187 Dresden,
Germany}
\author{Jœrg Sichelschmidt}
\affiliation{Max Planck Institute for Chemical Physics of Solids, 01187 Dresden,
Germany}
\author{Mark T.F. Telling}
\affiliation{ISIS Pulsed Neutron and Muon Source, STFC Rutherford Appleton Laboratory,
Harwell Campus, Didcot, Oxfordshire OX110QX, UK}
\author{P. K. Biswas}
\affiliation{ISIS Pulsed Neutron and Muon Source, STFC Rutherford Appleton Laboratory,
Harwell Campus, Didcot, Oxfordshire OX110QX, UK}
\author{M. Baenitz}
\affiliation{Max Planck Institute for Chemical Physics of Solids, 01187 Dresden,
Germany}
\author{I. Dasgupta}
\affiliation{School of Physical Sciences, Indian Association for the Cultivation
of Science, Jadavpur, Kolkata 700032, India}
\author{Sumiran Pujari}
\affiliation{Department of Physics, Indian Institute of Technology Bombay, Powai,
Mumbai 400076, India}
\author{A. V. Mahajan}
\email{mahajan@phy.iitb.ac.in}
\affiliation{Department of Physics, Indian Institute of Technology Bombay, Powai,
Mumbai 400076, India}
\date{\today}

\maketitle
Polycrystalline Sr$_{3}$CuSb$_{2}$O$_{9}$ (SCSO) sample was prepared by conventional solid state reaction
techniques with high purity ingredients. X-ray diffraction showed the presence of superlattice reflections. We have then followed up with magnetisation,
heat capacity, muon spin relaxation ($\mu$SR), electron spin resonance (ESR), and $^{121}$Sb
nuclear magnetic resonance (NMR) measurements. These are backed up
by density functional theory (DFT) calculations and predictions based on a Dirac spin liquid ansatz to build a comprehensive picture of the properties
of SCSO. We delve into various details in the following sections. 

\begin{figure}[h]
\centering{}\includegraphics[scale=0.28]{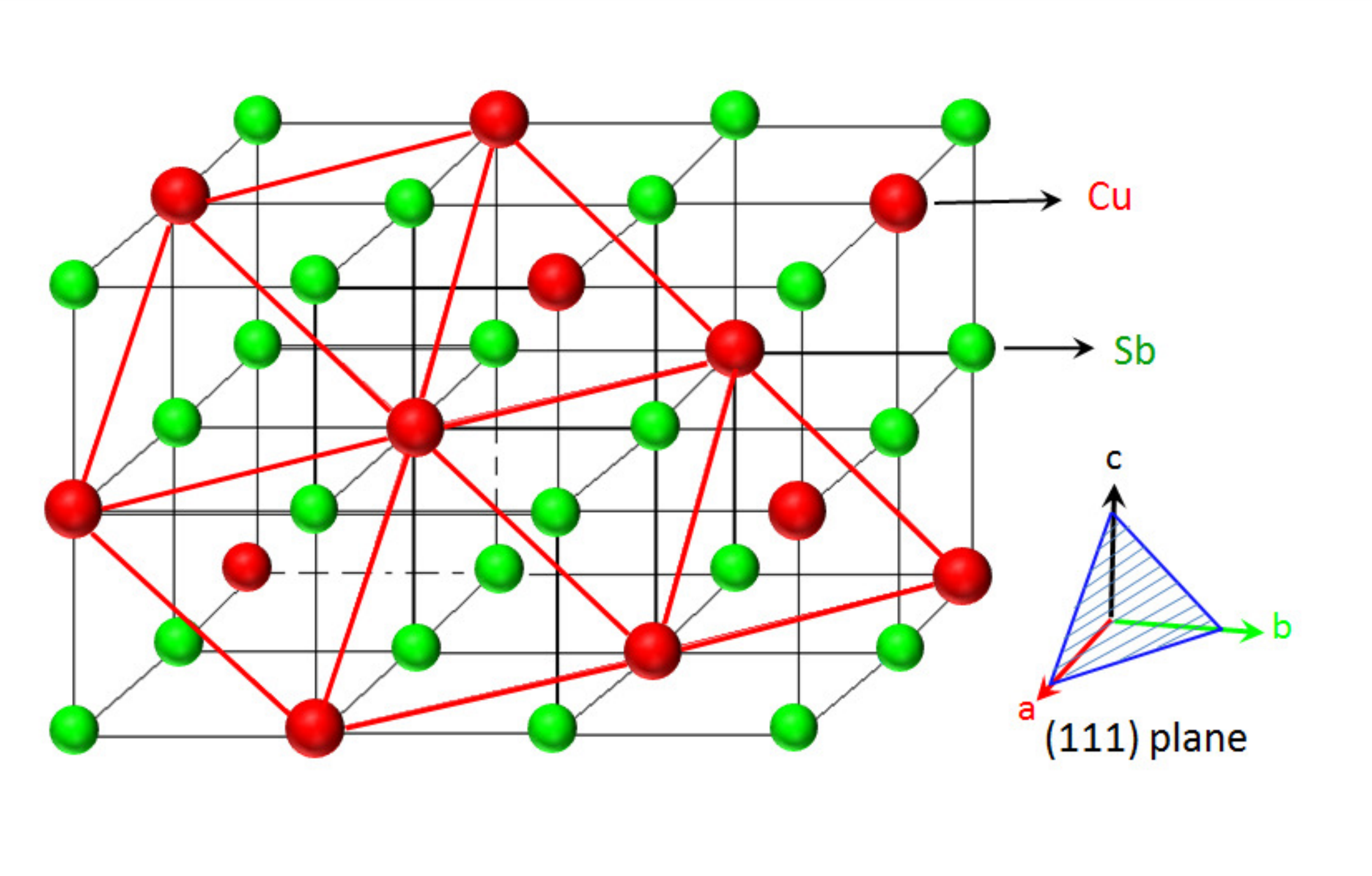}\caption{\label{fig:The-edge-shared-triangular _111 plane}The edge-shared
triangular lattice network of Cu$^{2+}$ atoms in the (111) plane of the
simple cubic lattice considering 1:2 cation ordering at the B-site
in the triple perovskite Sr$_{3}$Cu$\mathit{\mathrm{Sb}_{\mathrm{2}}}$O$_{9}$.}
\end{figure}

\section{X-ray diffraction and crystal structure}

The cubic perovskites are expressed by the general chemical formula
ABX$_{3}$. One will notice that the (111) planes present an edge-shared
triangular geometry. In real materials, in order to accommodate ions
of various radii, rotation and tilting of the BO$_{6}$ octahedra
takes place resulting in a lower than cubic symmetry for the crystal
structure. In the present case, the B site is occupied by Cu and Sb
ions. With 1:2 cation ordering at the B-site, the (111) planes in
the pseudocubic lattice will have successive Cu planes separated by
two Sb planes. These planes have an edge-shared triangular geometry
(shown in Fig. \ref{fig:The-edge-shared-triangular _111 plane}) which
is geometrically frustrated. 

\begin{figure}[h]
\centering{}\includegraphics[scale=0.35]{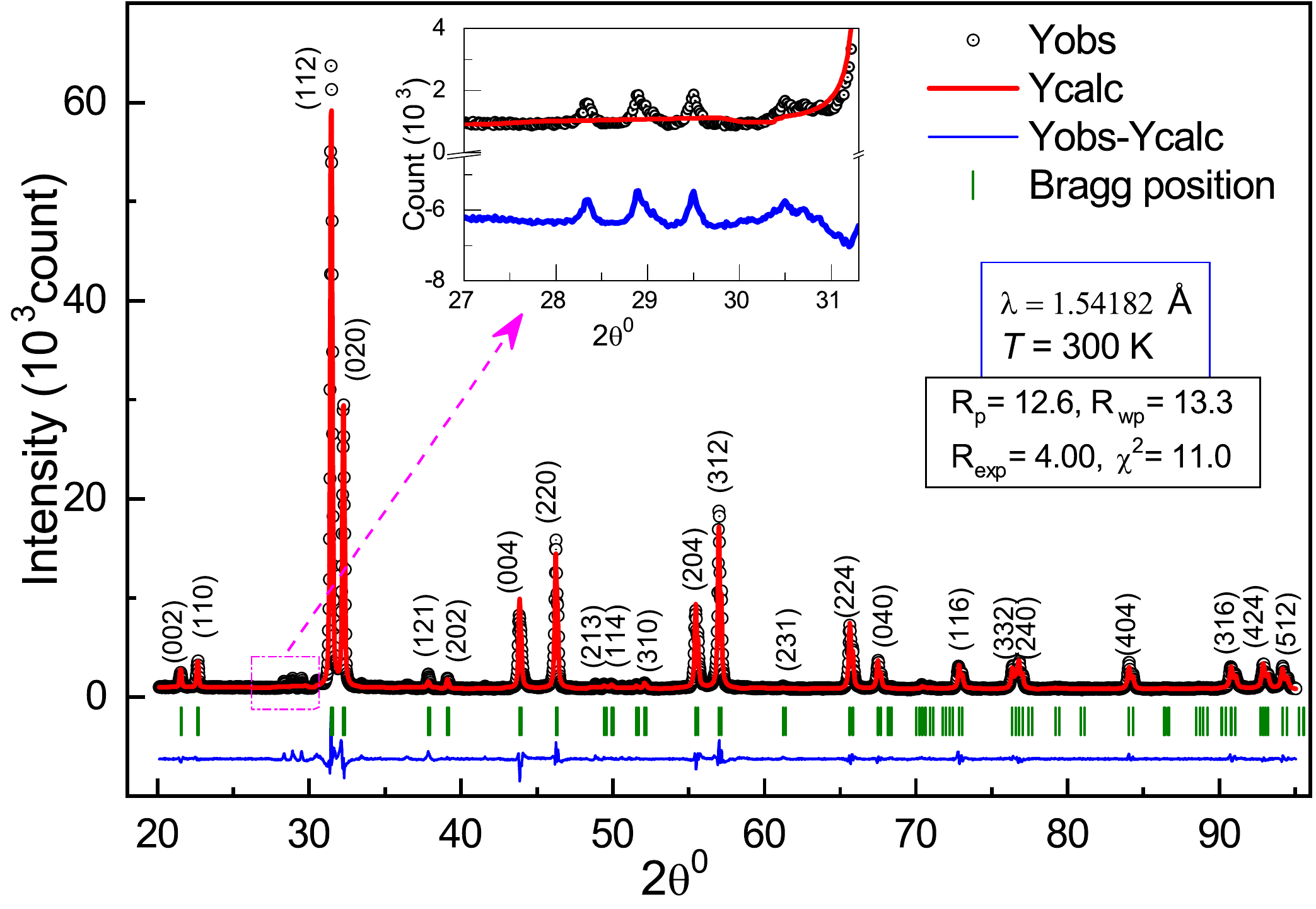}\caption{\label{fig:Powder-XRD-pattern SCSbO}{Rietveld refinement
of powder XRD pattern of Sr$_{3}$CuSb$_{2}$O$_{9}$ is shown along
with Bragg positions with corresponding Miller indices (hkl). Inset
shows the unindexed peaks at low angles which might be originating
from the superlattice reflections.}}
\end{figure}

\begin{table*}
	\centering{}\caption{\label{tab:Rietveld-parameter_SCSbO}{Obtained parameters after Rietveld refinement.}}
	\begin{tabular}{|lll|}
	
		\hline 
		Space group & $I{\rm 4}/mcm$ & $P\bar{1}$ \tabularnewline
		\hline 
		Propagation vector & None & $\vec{K}$ = ($\frac{1}{3}$, $\frac{1}{3}$, $\frac{1}{3}$) \tabularnewline
		Lattice parameter  & $\mathit{a}$ = $\mathit{b}$ = 5.547 \AA & $\mathit{a}$ = $\mathit{b}$ = 5.547 \AA \tabularnewline
		& $\mathit{c}$ = 8.248 \AA & $\mathit{c}$ = 8.248 \AA \tabularnewline
		$\alpha$, $\beta$, $\gamma$ & $\alpha=\beta=\gamma=  $90$^{\mathrm{o}}$ & $\alpha=\beta=\gamma= $90$^{\mathrm{o}}$ \tabularnewline
		Cell volume {[}\AA $^{3}${]} & 253.7844 & 253.7844 \tabularnewline
		R$_{\rm p}$, R$_{\rm wp}$, R$_{\rm exp}$(\%) & 12.6, 13.3, 4.00 & 9.76, 10.7, 4.50 \tabularnewline
		Bragg R-factor & 3.06 & 0.75 \tabularnewline
		RF-factor & 3.00 & 1.05 \tabularnewline
		$\chi^{2}$ & 11.0 & 5.62\tabularnewline
		\hline 
	\end{tabular}
\end{table*}

\begin{figure}[h]
\centering{}\includegraphics[scale=0.3]{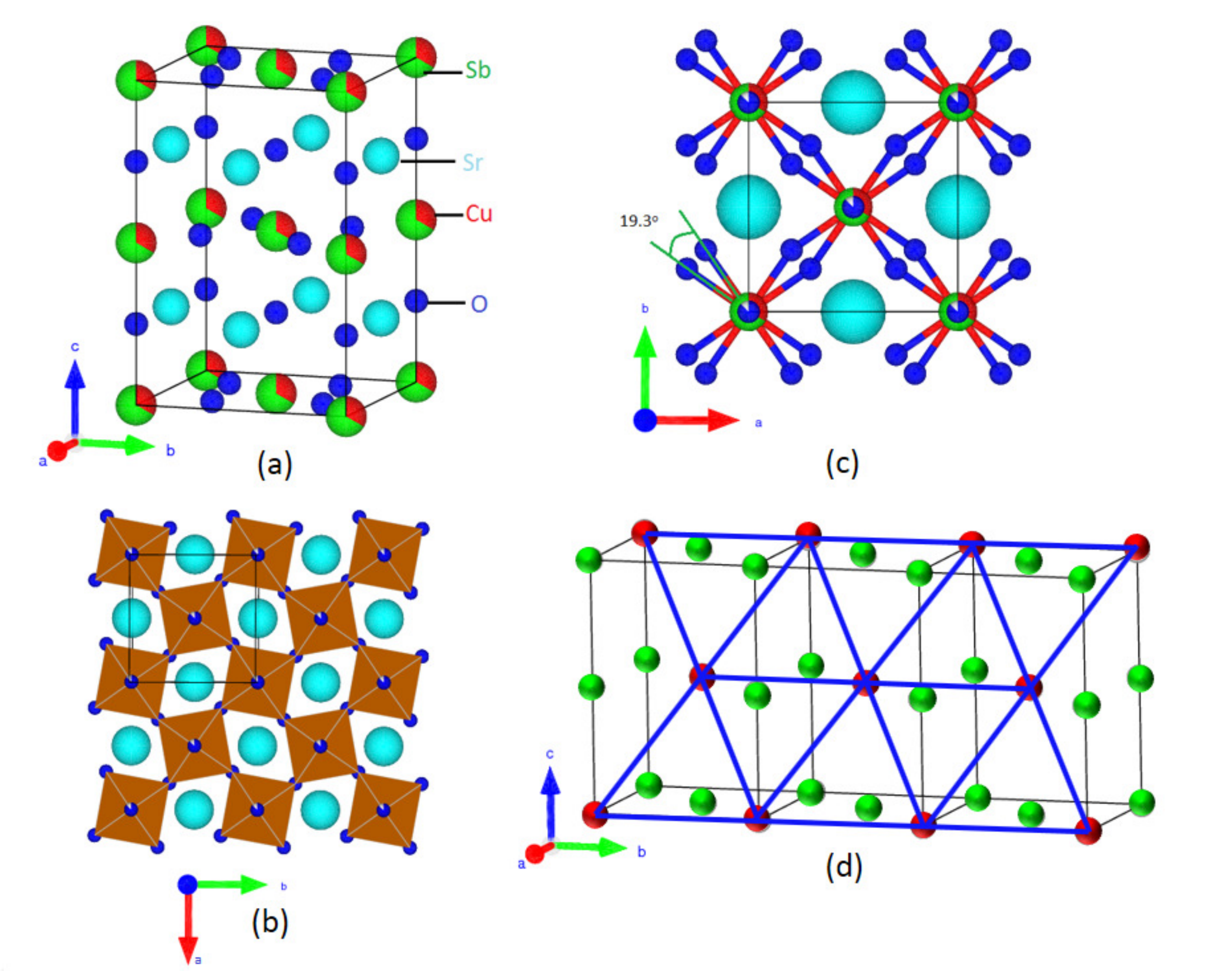}\caption{\label{fig:Structure-of-SCSbO}{Structure of Sr$_{3}$CuSb$_{2}$O$_{9}$
(a) One tetragonal unit cell showing the mixed B-site which is occupied
by both Sb$^{5+}$ and Cu$^{2+}$ ions, (b) Tilted corner-shared (Cu/Sb)
O$_{6}$ octahedra in $\mathit{ab}$-plane. (c) Top view of one unit
cell along $\mathit{c}$-axis, showing the tilting angle of 19.3$^{\mathrm{o}}$
between consecutive CuO$_{6}$ octahedra. (d) Edge-shared triangular
lattice network of Cu$^{2+}$ ions considering 1:2 cation ordering
at the B-site of the perovskite.}}
\end{figure}

The Rietveld refinement of Sr$_{3}$CuSb$_{2}$O$_{9}$ by Fullprof
suite software \citep{Carvajal1990}, in the body-centered tetragonal structure with space
group $I{\rm 4}/mcm$ (140), is shown in Fig. \ref{fig:Powder-XRD-pattern SCSbO}.  Atomic co-ordinates with site and occupancy of each element of Sr$_{3}$CuSb$_{2}$O$_{9}$
are given in Table \ref{tab:Atomic-positions-in SCSbO}. The goodness
of the Rietveld refinement is defined by the following parameters:\textbf{
}${\normalcolor \chi^{2}}$ = 11.0; ${\normalcolor R}_{{\normalcolor \mathrm{p}}}$=
12.6\%; ${\normalcolor R}_{{\normalcolor \mathrm{wp}}}$= 13.3\%;
${\normalcolor R}_{\mathrm{{\normalcolor exp}}}$= 4.00\%. 
A few
unindexed (possibly superlattice) peaks at low angles (2$\theta$ < $30^{\mathrm{o}}$) are shown
in the inset of Fig. \ref{fig:Powder-XRD-pattern SCSbO}.  To determine the possible propagation vector corresponding to these superlattice peaks, we performed a propagation vector ($\vec{K}$) search and fitting. During this $\vec{K}$-search and $\vec{K}$-vector fitting, the lattice parameters were fixed to the refined values of the tetragonal lattice with space group $I{\rm 4}/mcm$. However, a lower symmetry spacegroup $P\bar{1}$ (triclinic) was used to allow all possible superlattice reflections.  The  LeBail fit was employed for the $\vec{K}$-search.
We could then index the additional reflections as superlattice reflections with $\vec{K}$ = (1/3, 1/3, 1/3)  and index the full pattern.  Note that the structure of SCSO is quite different from the various (either with the triangular or honeycomb Cu-planes) structures of Ba$_3$CuSb$_2$O$_9$ (BCSO). In BCSO, the basal $a-b$ plane has a triangular magnetic lattice whereas for SCSO it is the (111) planes which present the triangular lattice. Another illustration of the crystal structure (using the Vesta software \citep{Momma2011}) is shown in Fig. \ref{fig:Structure-of-SCSbO}.
\begin{table}[H]
\centering{}\caption{\label{tab:Atomic-positions-in SCSbO}{Atomic coordinates
in Sr$_{3}$CuSb$_{2}$O$_{9}$ based on a $I{\rm 4}/mcm$ space group refinement.}}
\vspace{0.5cm}
\begin{tabular}{cccccc}
\hline 
Atom & Wyckoff position & x & y & z & Occupancy\tabularnewline
\hline 
\hline 
Sr & 4b & 0.000 & 0.500 & 0.250 & 1.000\tabularnewline
Cu & 4c & 0.000 & 0.000 & 0.000 & 0.333\tabularnewline
Sb & 4c & 0.000 & 0.000 & 0.000 & 0.667\tabularnewline
O1 & 8h & 0.794 & 0.294 & 0.000 & 1.000\tabularnewline
O2 & 4a & 0.000 & 0.000 & 0.250 & 1.000\tabularnewline
\hline 
\end{tabular}
\end{table}
\section{Magnetization}
We measured the dc magnetization $\mathit{M}(T)$ as a function of
temperature in the  range 2-300 K on a hard pellet of $\mathrm{SCSO}$
in zero field cooled (ZFC) and field cooled (FC) mode at several magnetic
fields ($\mathit{H}$) using a Quantum Design MPMS system. The main
features of our observations from the measurements are described below.
\begin{figure}[h]
\centering{}\includegraphics[scale=0.30]{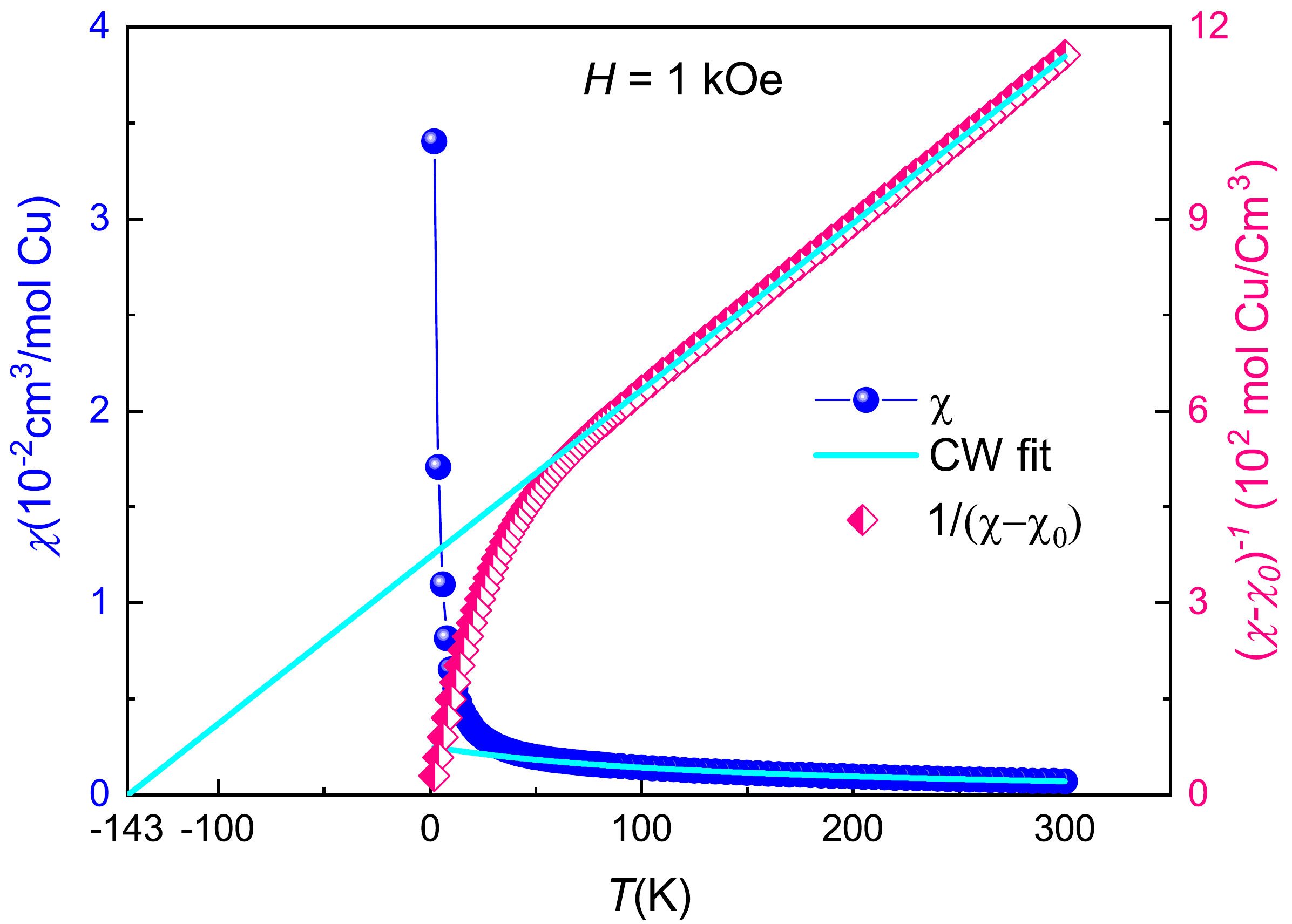}\caption{\label{fig:Curie-Weiss-fit-of SCSbO}{The left $\mathit{y}$-axis
shows the temperature dependence of $\mathit{\chi}(T)=\frac{M(T)}{H}$
(blue circles) of SCSO measured in $H$ = 1 kOe and the right $\mathit{y}$-axis
shows the inverse susceptibility free from the temperature independent
susceptibility $\chi_{0}$. The Curie-Weiss (CW) fit is shown in the
temperature range 150 - 300 K with a cyan solid line. The intercept
on the $\mathit{x}$-axis gives a CW temperature $\theta_{\rm CW}\simeq$
-143 K.}}
\end{figure}
The susceptibility of SCSO (Fig. \ref{fig:Curie-Weiss-fit-of SCSbO})
was analysed initially using $\chi=\chi_{0}+\frac{C}{T-\theta_{\rm CW}}$
; where $\chi_{0}$, $C$ and $\theta_{\rm CW}$ are temperature independent
susceptibility (arising from the diamagnetic core ($\chi_{\rm core}$)
and paramagnetic van-Vleck ($\chi_{\rm VV}$) contributions), Curie constant,
and the Curie-Weiss temperature, respectively. A fit in the range
of 150 - 300\,K gave $\chi_{0}$= -1.61$\times$ 10$^{-4}$ (cm$^{3}$/mol
Cu), $\mathit{C}$ = 0.384\,(Kcm$^{3}$/mol Cu) and the $\theta_{\mathrm{CW}}
= -143 $ K. From the Curie constant, we obtained the effective moment
$\mu_{\rm eff}$ = 1.75\,$\mu_{\rm B}$ which is close to the expected value
of $\mu_{\rm eff}$ = 1.73\,$\mu_{\rm B}$ for $\mathit{S}$ = $\frac{1}{2}$
Cu$^{2+}$ ion. The negative value of CW temperature indicates antiferromagnetic
correlations between the Cu$^{2+}$ magnetic ions. Given that there
is no order down to 65 mK (see $\mu$SR analysis in a latter section),
this gives a very high frustration parameter ($\mathit{f}$ = $\frac{\theta_{\rm CW}}{T_{\rm N}}$
>2000). 
\begin{figure}[h]
\centering{}\includegraphics[scale=0.30]{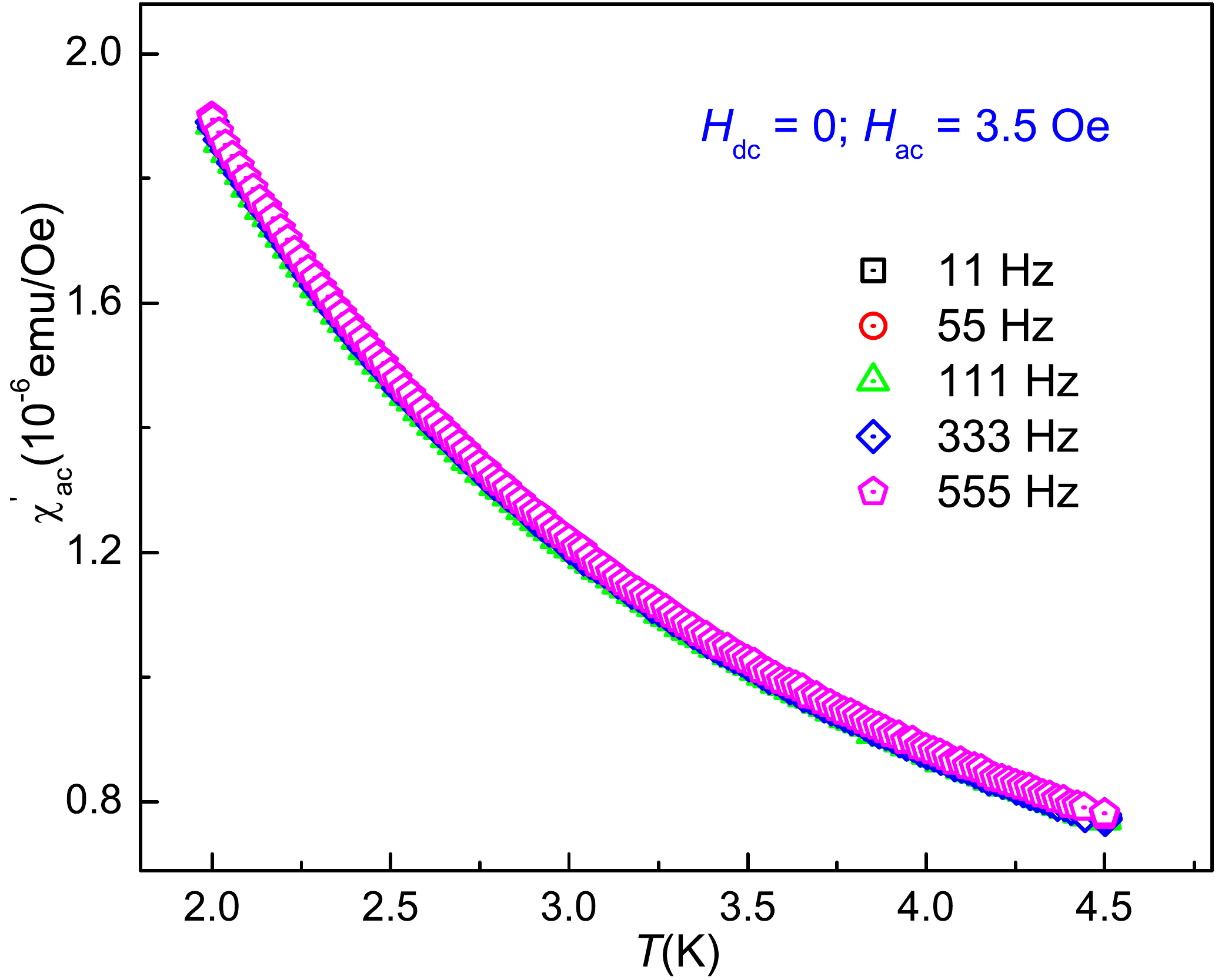}\caption{\label{fig:M_H and ac susceptibility SCSbO} {The in phase component
of the ac susceptibility of SCSO as a function of temperature at various frequencies.}}
\end{figure}
The ac susceptibility of SCSO was measured to ensure that there is
no glassy behaviour at low temperature. Fig. \ref{fig:M_H and ac susceptibility SCSbO}
shows the in-phase component of a.c. susceptibility as a function
of temperature at different frequencies with an a.c. field of 3.5
Oe amplitude and zero d.c. applied field. No anomaly is seen excluding the presence of a glassy phase down to 1.8 K.

\section{Electron Spin Resonance}
Electron spin resonance (ESR) is another microscopic probe for magnetism.
ESR can probe the spin dynamics of the very spin system in contrast
to NQR and \textgreek{m}SR, both of which access the spin system indirectly.
Thus, ESR has proven to provide direct access to the spin dynamics
of many systems. Here, the ESR experiments probe the Cu$^{2+}$ spins
of SCSO at X-band frequencies ($\nu$ = 9.4 GHz) using a continuous-wave
ESR spectrometer. The sample was measured as a powder embedded in paraffine and the temperature was set with a helium-flow
cryostat allowing for temperatures between 3 K and 290 K. 

\begin{figure}[h]
\centering{}\includegraphics[scale=0.35]{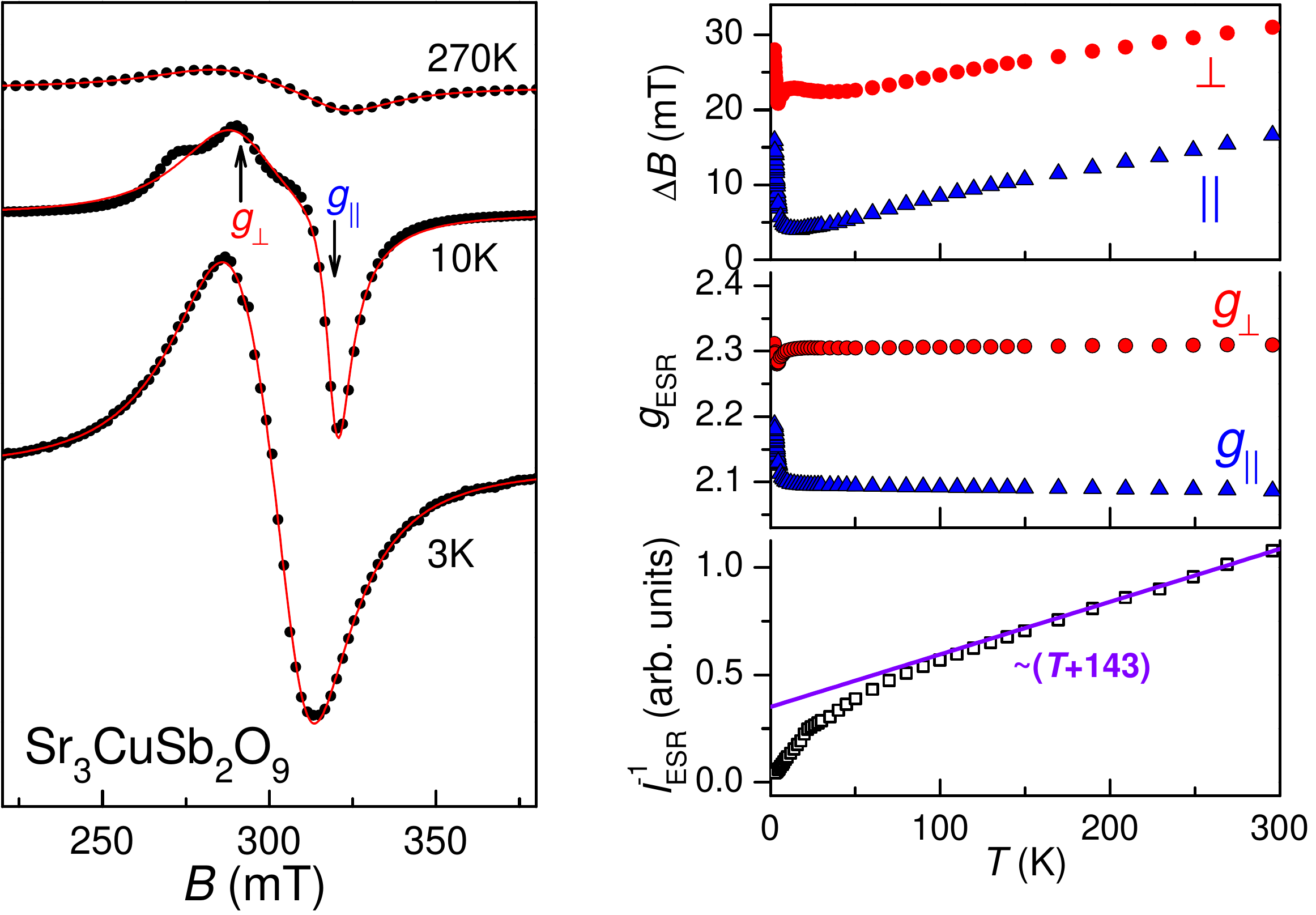}\caption{\label{fig:ESR_SCSbO}{Left panel: X-band ESR spectra (symbols) at representative temperatures and fitted uniaxial powder-averaged Lorentzians (solid lines). Fitted values of anisotropic g-values are indicated for the spectrum at $\mathit{T}$ = 10 K. Right panel: Temperature dependence of fitted parameters linewidth, effective g-value, and inverse integrated ESR intensity with Curie-Weiss behavior as indicated by the solid line.}}
\end{figure}

Typical ESR spectra (given as the first derivative of the absorbed microwave power) of SCSO are shown in the left panel of Fig. \ref{fig:ESR_SCSbO} for selected temperatures. The lines could reasonably be fitted by a Lorentzian shape, averaged for uniaxial g-factor anisotropy, as shown by the solid lines. This yield the linewidth ${\Delta B}$ which is a measure of the
spin-probe relaxation rate, and the resonance field ${B_{\rm res}}$
which is determined by the effective g-factor (${g_{\rm ESR}=h\nu/\mu_{\rm B}B_{\rm res}}$)
and internal fields. 
Below $T\simeq$  80 K the spectra reveal additional lines which indicate the presence of Cu$^{2+}$ spins in a lower than uniaxial environmental symmetry. These results are similar to those in Ba$_3$CuSb$_2$O$_9$ \cite{Han2015}. At higher temperatures and below about 5 K these structures becomes smeared out by line broadening. As shown in the right panel of Fig. \ref{fig:ESR_SCSbO} both the linewidth $\Delta B$ and $g_{\rm ESR}$ show a considerable temperature dependence below $\mathit{T} \simeq$ 10 K, indicating emergent spin fluctuations and internal magnetic fields. For $\mathit{T}$ = 290 K we obtained g$_{\parallel}$ = 2.08 and g$_{\perp}$
= 2.31, corresponding to an averaged value $g_{\rm avg}=\sqrt{(g_{\parallel}^{2}+2g_{\perp}^{2})/3}$ = 2.24 which
is typical of Cu$^{2+}$ for many cuprate systems. 
The ESR intensity $\mathit{I_{\rm ESR}}$
corresponds to the integrated ESR absorption and is determined by the static spin-probe susceptibility, $\mathit{i.e.}$ it provides
a direct microscopic probe of the sample magnetization. The right panel of Fig. \ref{fig:ESR_SCSbO} also shows the temperature dependence of  $I^{-1}_{\rm ESR}$ which follows the measured susceptibility  very well with a Weiss temperature of  $-143$~K  for the high temperature behavior. 

\section{Heat Capacity}

\begin{figure}[h]
	\centering{}\includegraphics[scale=0.3]{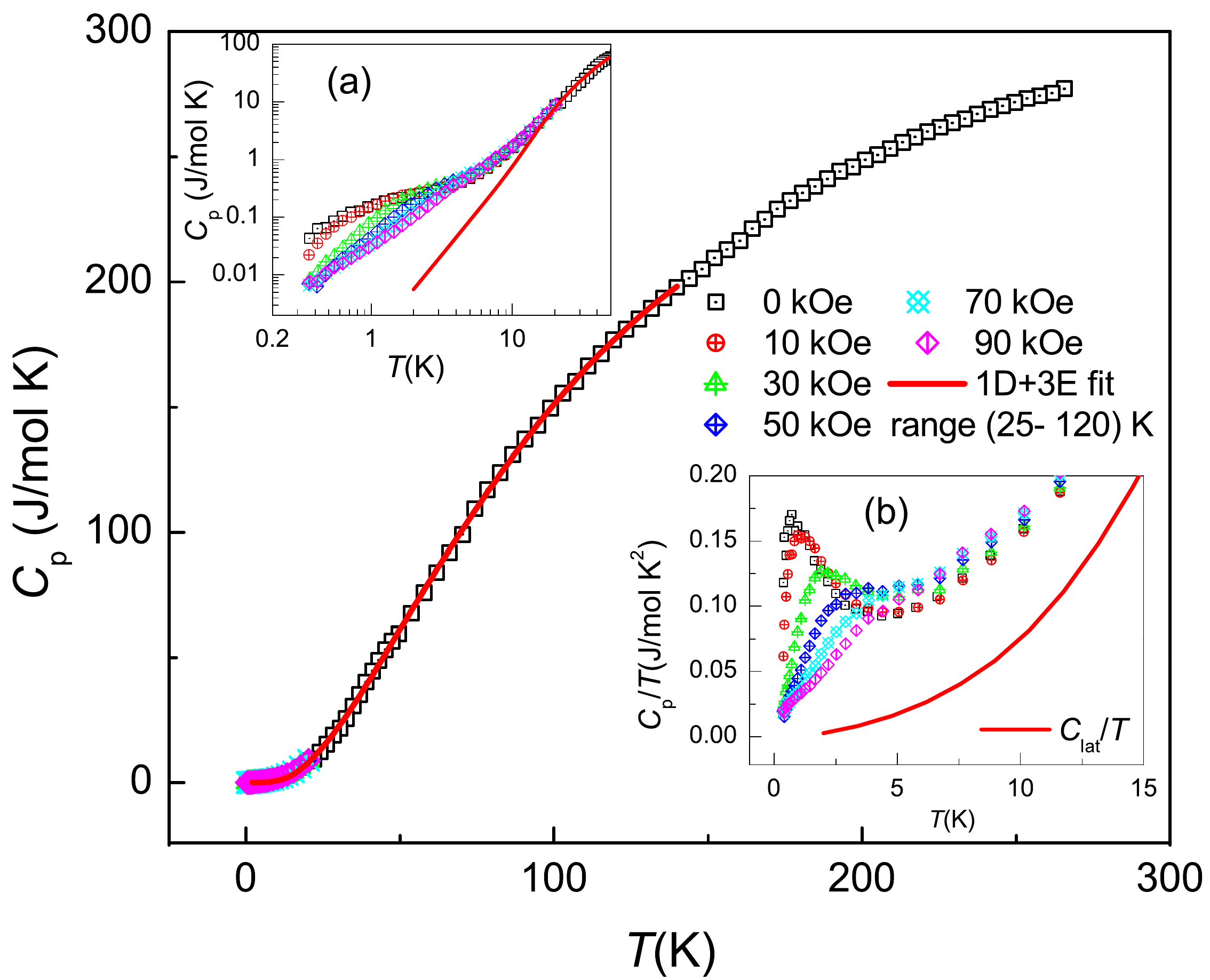}\caption{\label{fig:HC_of SCSbO}{Heat capacity of SCSO at different
			fields with Debye plus Einstein fitting. Inset (a) shows the same data on a log-log scale. In inset (b),  $\mathit{C_{\mathrm{p}}}$/$\mathit{T}$
			vs. $\mathit{T}$ shows low temperature anomaly due to Schottky effect.}}
\end{figure}

\begin{figure}[H]
	\centering
	\includegraphics[width=0.8\linewidth]{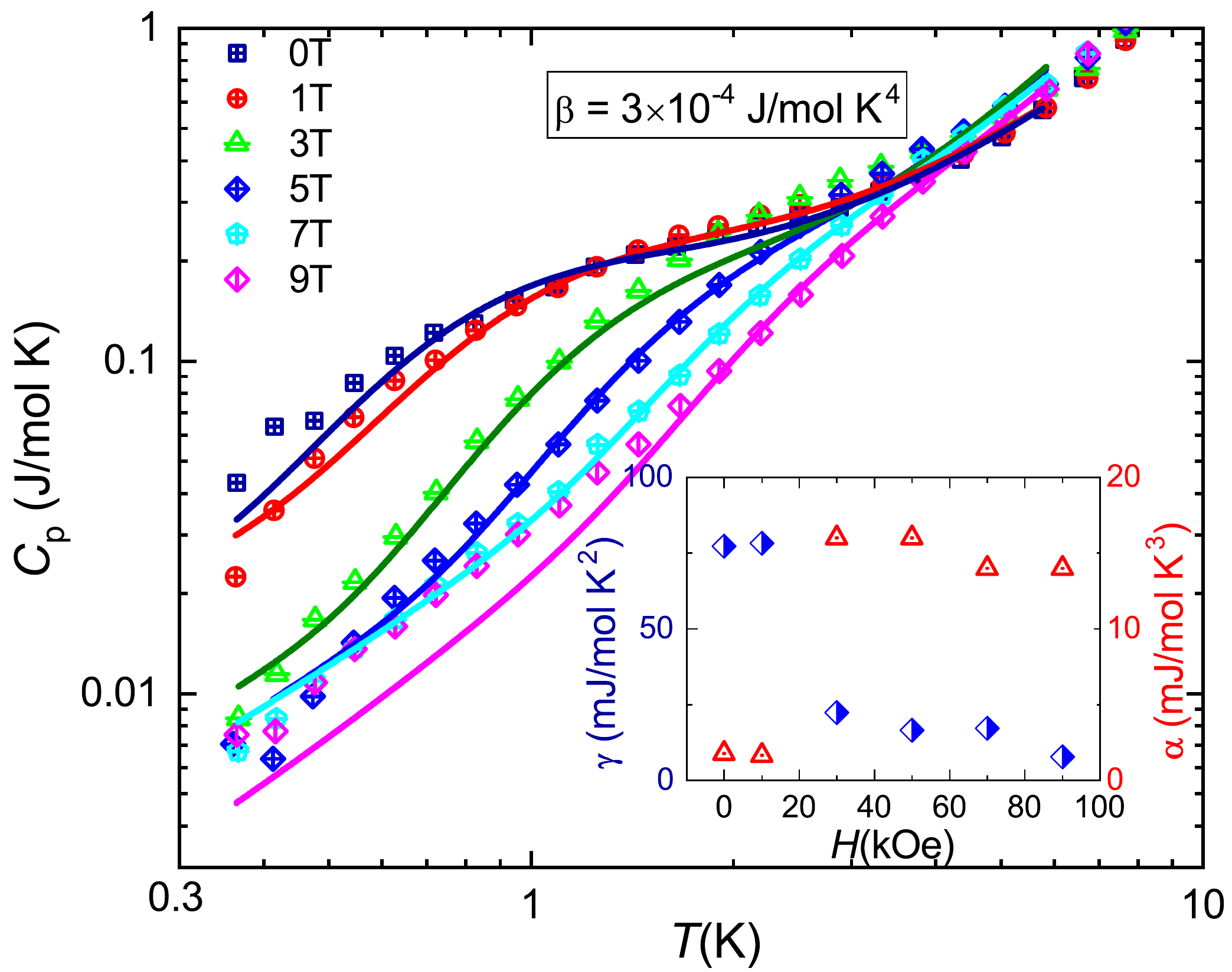}
	\caption{The specific heat of SCSO is shown together with fits (solid lines) as described in the text. The inset shows the variation of $\gamma$ and $\alpha$ with $H$.}
	\label{fig:fig4newb}
\end{figure}

\begin{figure}[H]
	\centering
	\includegraphics[width=0.8\linewidth]{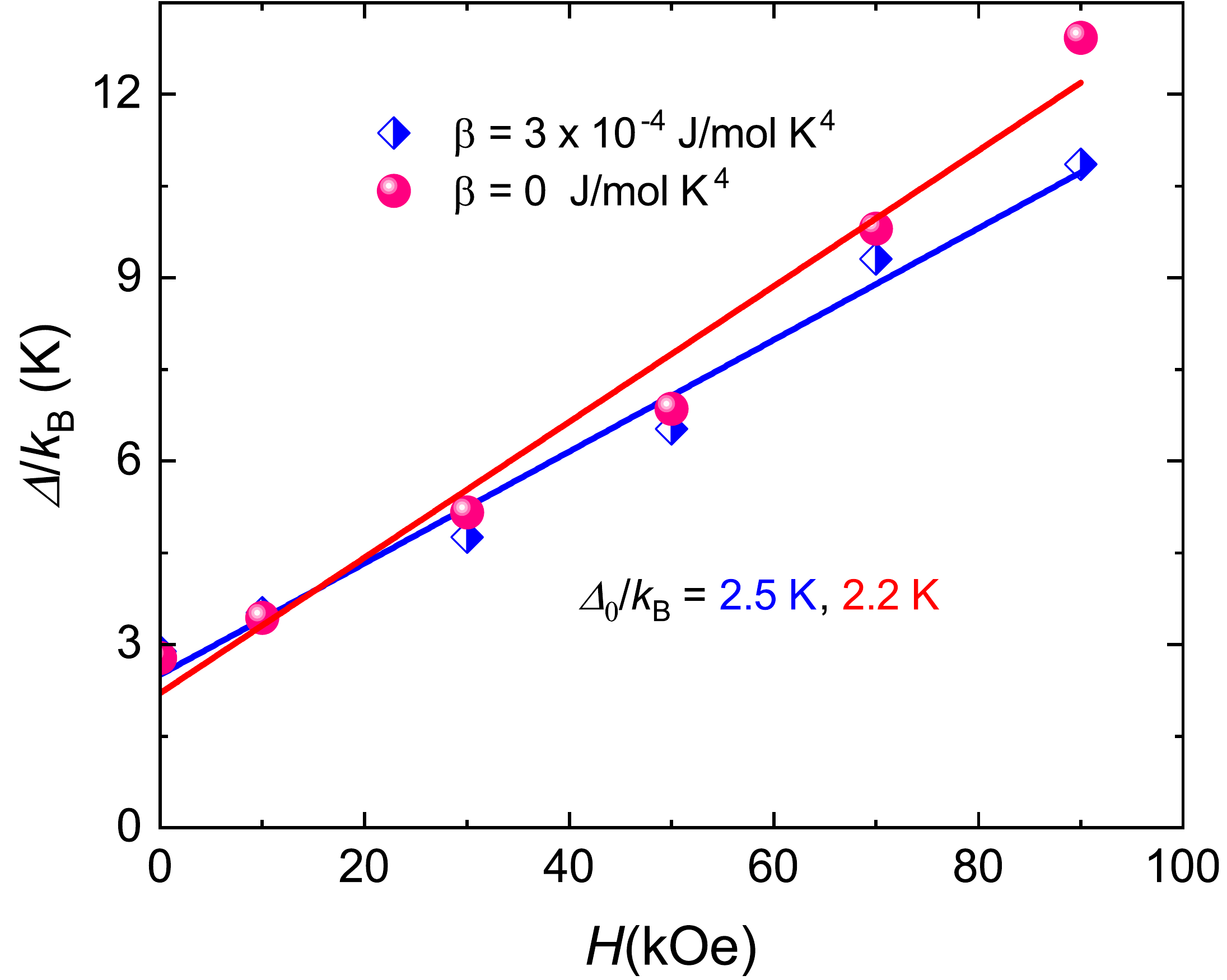}
	\caption[Delta/kB]{Linear variation and fit of Schottky gap for both values of $\beta$. They show intrinsic interaction present in the system as in 0 kOe field we get a non-zero intercept close to 2.5 K.}
	\label{fig:schottky-gap-scso}
\end{figure}

Fig. \ref{fig:HC_of SCSbO}
shows the $\mathit{C_{\mathrm{p}}(T)}$ vs. $\mathit{T}$ plot in
different fields. For clarity, we have shown $\mathit{C_{\mathrm{p}}(T)}$
vs. $\mathit{T}$ plot in log-log scale in the inset (a) of Fig. \ref{fig:HC_of SCSbO}.
Also, we have plotted $\mathit{C_{\mathrm{p}}(T)/T}$ vs. $\mathit{T}$
in the inset (b) of Fig. \ref{fig:HC_of SCSbO}, which indicates the
presence of the Schottky anomaly below 9\,K due to free spins within
the system as the peak position is shifted gradually towards high
temperature as field strength increases. 
Since we do not have a suitable non-magnetic analog of this system, to obtain the lattice contribution, 
we fitted the heat capacity in zero field in the high-$T$ region (such as 50 -130 K) to a combination of one Debye
and three Einstein (1D+3E) functions (see later). This was  then extrapolated to lower 
temperatures for analysing the low-$T$ data. A $\beta T^3$ variation of the lattice heat capacity was seen (for $T < 6$ K) from this extrapolation. The value of $\beta$ depends somewhat on the high-$T$ fitting range used for the lattice fit but is around 3 $\times$ 10$^{-4}$ J/mol K$^4$.
  Using this value good fits of the measured data to $C_{\mathrm{p}}(T) = \gamma T + \alpha T^2 + \beta T^3 + fC_\mathrm{Schottky}$ are obtained (see Fig.\ref{fig:fig4newb}).  The variation of the inferred Schottky gap with the applied field  is shown in Fig. \ref{fig:schottky-gap-scso}.
    
We used a combination of Debye and Einstein terms as expressed by the Eq.
\ref{eq:Debye} and \ref{eq:Einstein} to estimate the lattice contribution $\mathit{C_{\mathrm{lat}}}$.
Here, $\mathit{C_{\mathrm{D}}}$ and $\mathit{C_{\mathrm{E_{i}}}}$ are
the weightage factors corresponding to acoustic and optical modes
of atomic vibrations and $\theta_{\mathrm{D}}$, $\theta_{\mathrm{E}_{\mathrm{i}}}$
are the corresponding Debye and Einstein temperatures, respectively.

\begin{equation}
C_{\mathrm{Debye}}(T)=C_{\mathrm{D}}\left[9R(\frac{T}{\theta_{\mathrm{D}}})^{3}\intop_{0}^{x_{\mathrm{D}}}\frac{x^{4}e^{x}}{(e^{x}-1)^{2}}dx\right]\label{eq:Debye}
\end{equation}
\begin{equation}
C_{\mathrm{Einstein}}(T)=\sum C_{\mathrm{E_{i}}}\left[3R(\frac{\theta_{\mathrm{E}_{i}}}{T})^{2}\frac{exp(\frac{\theta_{\mathrm{E}_{i}}}{T})}{(exp(\frac{\mathrm{\theta}_{\mathrm{E}_{i}}}{T})-1)^{2}}\right]\label{eq:Einstein}
\end{equation}

We obtained the weightage factors in the ratio: $C_{\mathrm{D}}$:$C_{\mathrm{E_{1}}}$:$C_{\mathrm{E_{2}}}$:$C_{\mathrm{E_{3}}}$
= 1:1:5:6. The total sum of $C_{\mathrm{D}}$+$\sum C_{\rm E_{i}}$ is
equal to 13 which is close to the total number of atoms (15) per formula
unit of SCSO. The fitting also yields the Debye temperature $\theta_{\rm D}$=
(140$\pm1)$\,K and the Einstein temperatures; $\theta_{\rm E_{1}}$=
(109$\pm1$)\,K, $\theta_{\rm E_{2}}$= (219$\pm2$)\,K and $\theta_{\rm E_{3}}$=
(436$\pm2$)\,K.

In Fig.\ref{fig:fig4newb}, we have fitted the $C_{p}$ vs. $\mathit{T}$ by
Eq. \ref{eq:HC fit}.

\begin{equation}
C_{p}=\gamma T+\alpha T^{2}+\beta T^{3}+fC_{\rm Sch}\label{eq:HC fit}
\end{equation}

where $\mathit{C_{\rm Sch}}$ is for two level system with $\mathit{S}$
=1/2.

\begin{equation}
C_{\rm Sch}=\left[R(\frac{\Delta}{k_{\rm B}T})^{2}\frac{exp(\frac{\Delta}{k_{\rm B}T})}{[1+exp(\frac{\Delta}{k_{\rm B}T})]^{2}}\right]\label{eq:schottky}
\end{equation}

Here $\mathit{f}$ is the fraction of free spins within the system,
$\Delta$ is the Schottky gap, $\mathit{R}$ is the universal gas
constant, $k_{\rm B}$ is the Boltzmann constant.
\begin{figure}[H]
	\centering
	\includegraphics[width=0.8\linewidth]{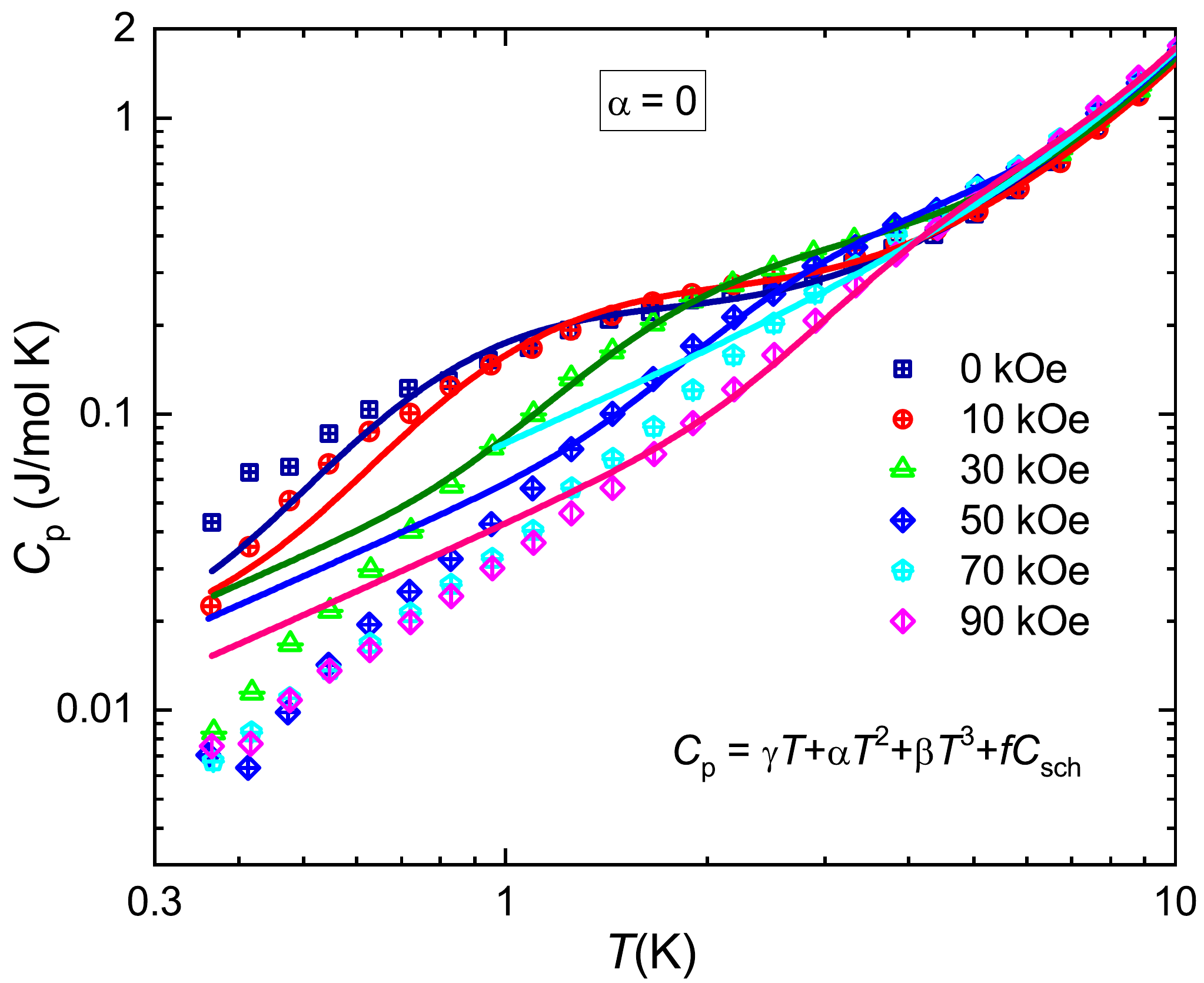}
	\caption[alpha=0]{The specific heat data of SCSO for all fields are shown with fits for $\alpha$ = 0 as mentioned in the text.}
	\label{fig:scsohcalpa0}
\end{figure}

\section{Nuclear Magnetic Resonance}

NMR is a useful probe of low energy excitations in magnetic insulators. 
For a recent example of a NMR study on a triangular system, see Ref. \cite{Zeng2020}. 
In our system, the $^{121}$Sb nucleus has a nuclear spin of $\mathit{I}$ = $\frac{5}{2}$,
natural abundance is 57.25\% and $\frac{\gamma}{2\pi}$ = 10.19 MHz/Tesla.
In our 94\,kOe fixed field NMR set up, we have initially checked
the spectra at different temperatures. As it was broader than 4\,MHz
 at 80\,K it was not possible to get the full NMR line shape
by single Fourier transform of the spin echo. Also the $^{121}$Sb NMR signal
is very weak at high temperature, and above 50 K it really requires
a large number of scans for a reasonable signal to noise ratio which
is very much time-consuming. We then measured $^{121}$Sb NMR spectra
by sweeping magnetic field at temperatures below $\mathit{T}$ = 50 K.

\subsubsection{$^{121}$Sb NMR Spectra}

\begin{figure}[h]
\centering{}\includegraphics[scale=0.3]{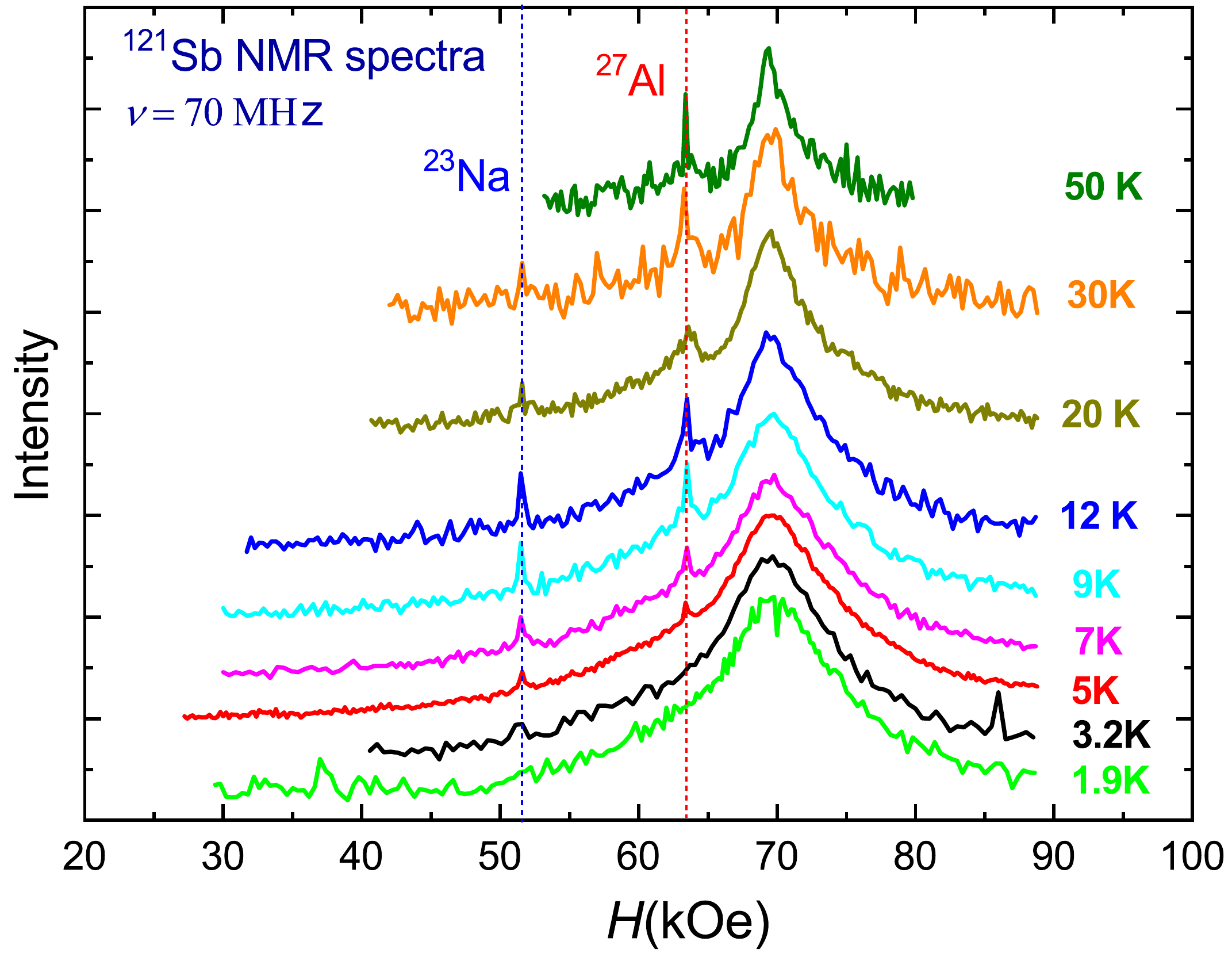}\caption{\label{fig:All-Sb Spectra SCSBO} {Field swept $^{121}$Sb
NMR spectra of Sr$_{3}$CuSb$_{2}$O$_{9}$ at various temperatures.}}
\end{figure}

\begin{figure}[h]
\centering{}\includegraphics[scale=0.3]{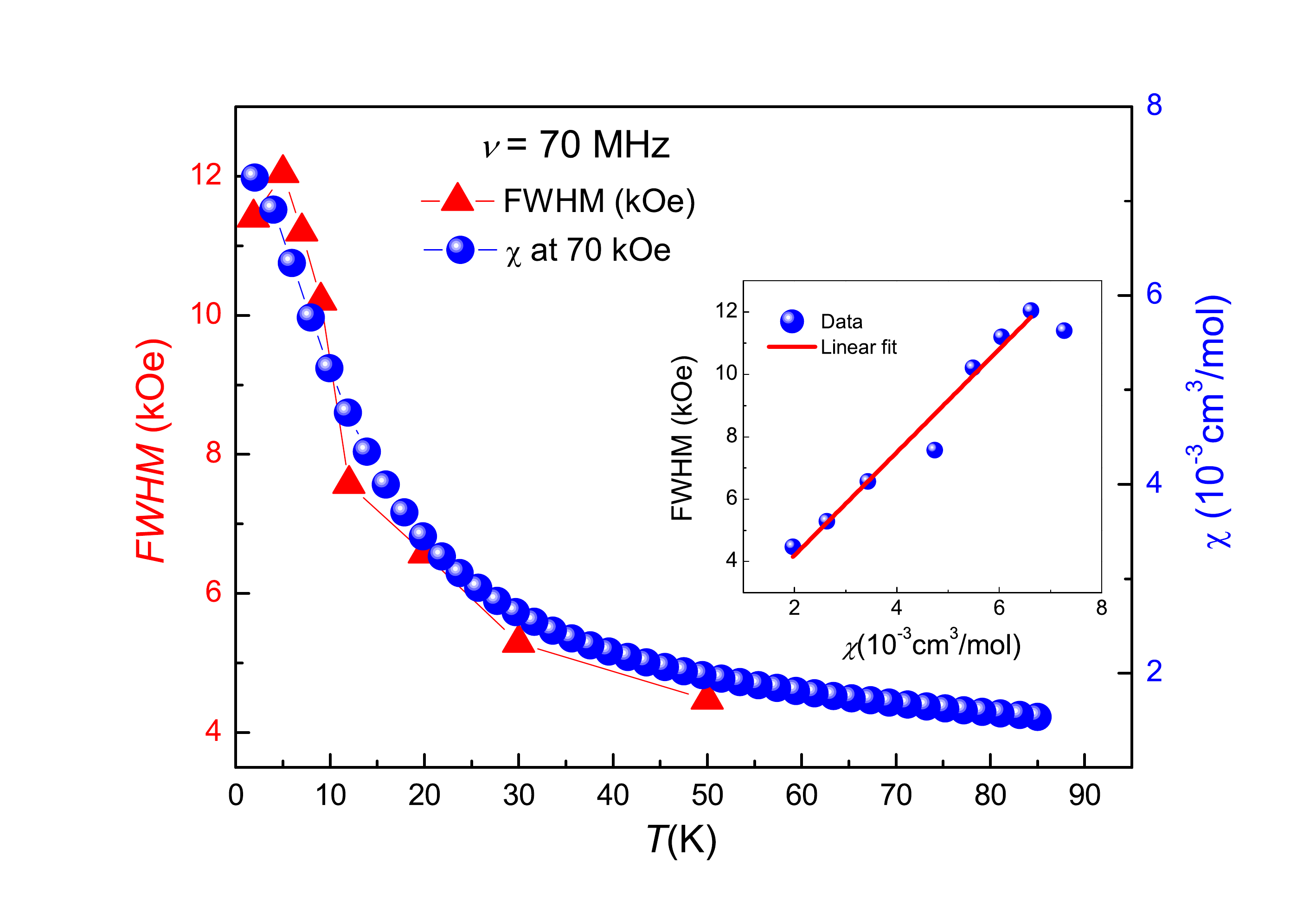}\caption{ \label{fig:-FWHM-as SCSbO}{FWHM of $^{121}$Sb NMR line (left $y$-axis) in SCSO is shown as a function of temperature together with the dc susceptibility at 70\,kOe (right $y$-axis). The inset shows that the FWHM scales with the susceptibility with temperature as an implicit parameter.}}
\end{figure}

We have measured field swept NMR spectra of $^{121}$Sb nucleus from
50\,K to 2\,K (shown in Fig.
\ref{fig:All-Sb Spectra SCSBO}) at a fixed frequency $\nu=$ 70 MHz. From the plot, it is clear that
there is no significant NMR line shift of the peak position as temperature
decreases. The spectrum broadens as temperature decreases. This broadening
is likely due to $^{121}$Sb nuclear-Cu$^{2+}$ local moment dipolar interaction. In each spectrum,
there are spikes at 63.4\,kOe and 51.6 kOe. These do not shift with
temperature and may be from the probe head or the sample holder. To
be sure, we performed measurements  under the same conditions after removing the
sample from the sample holder. Then we noticed that these signals
were present without sample also. It appears that the signals at 63.4
kOe and 51.6 kOe are from $^{27}$Al and $^{23}$Na nuclei which are
present in the probe head. The FWHM of the $^{121}$Sb NMR line shape
is plotted with temperature and it follows the bulk dc susceptibility
at 70\,kOe (shown in Fig. \ref{fig:-FWHM-as SCSbO}). The inset of
Fig. \ref{fig:-FWHM-as SCSbO} shows the linearity of FWHM with bulk
susceptibility data at 70 kOe.
\subsubsection{Spin-lattice relaxation}

\begin{figure}[h]
\centering{}\includegraphics[scale=0.25]{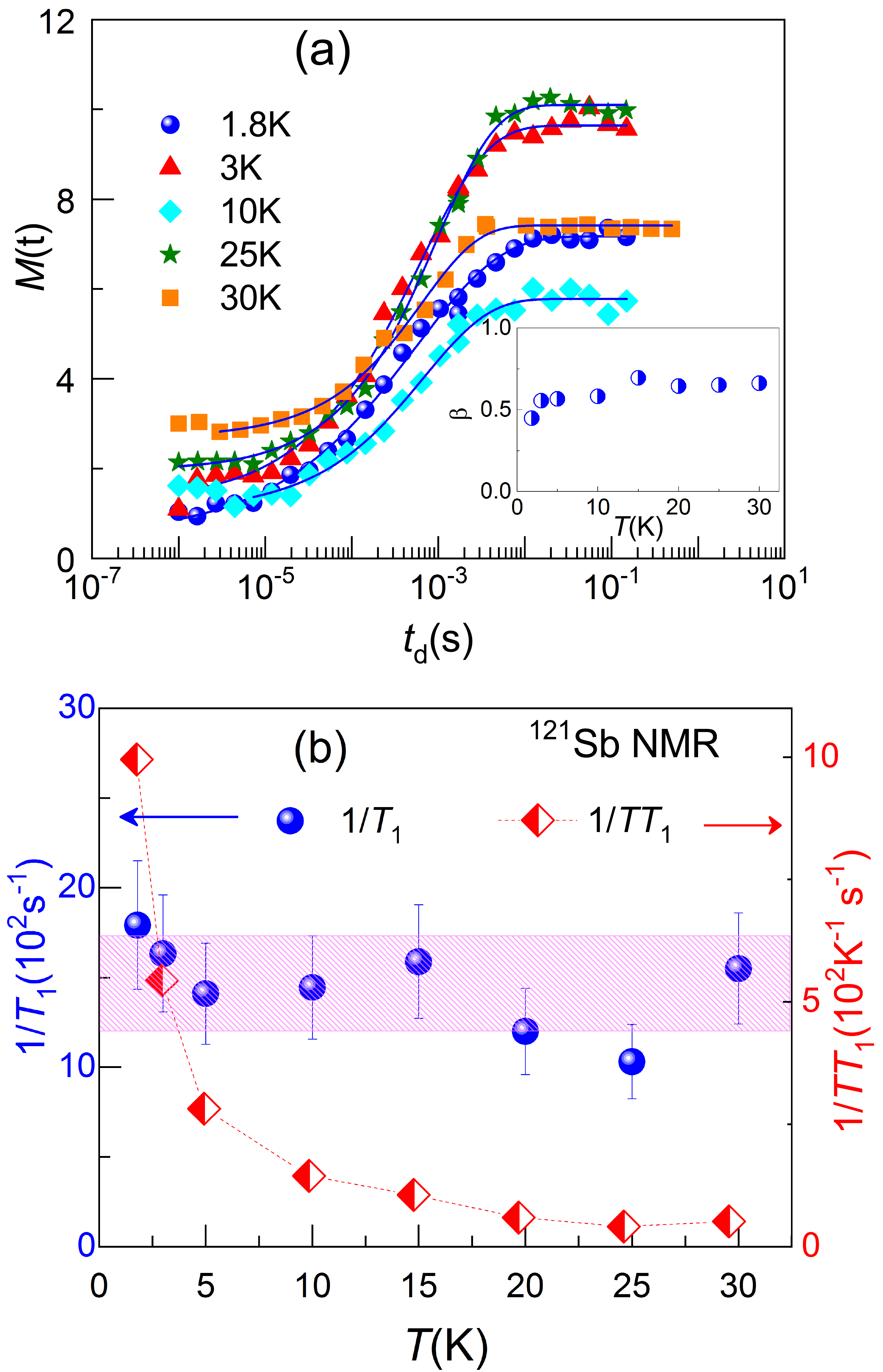}\caption{\label{fig:Saturation-recovery-of SCSBO}{Recovery of the
longitudinal nuclear magnetization for $^{121}$Sb as a function of
delay time after a saturating pulse sequence. The solid lines show
fits to stretched exponentials. Inset shows the variation of the stretching
exponent $\beta.$ (b) The spin-lattice relaxation rate (1/$\mathit{T_{1}}$)
is seen to be nearly independent of temperature. 1/$\mathit{T}$$_{1}$$\mathit{T}$
varies in a  Curie-Weiss-like manner. The dashed line is a guide to the eye.}}
\end{figure}

Fig. \ref{fig:Saturation-recovery-of SCSBO}(a) shows the saturation
recovery of the longitudinal component of $^{121}$Sb nuclear magnetization
$\mathit{i.e.}$ the spin-lattice relaxation ($\mathit{T}_{1}$)
from 30\,K to 2\,K. Using a pulse comb to saturate the broad line, we could  obtain about 80\% saturation of the signal. The nuclear magnetization recovery is well fitted with a stretched exponential
($M(t)=M_{0}[1-A\thinspace exp(\frac{-t}{T_{1}})^{\beta}]$).  The stretching exponent is around
$\beta$ $\approx$ 0.60 (shown in the inset of Fig. \ref{fig:Saturation-recovery-of SCSBO}(a)).
There is no significant change in values of $\mathit{T_{1}}$ with
temperatures and the average $\mathit{T_{1}}$ is around 650\,$\mu$s.
The spin-lattice relaxation rate is shown in Fig. \ref{fig:Saturation-recovery-of SCSBO}(b).
1/$\mathit{T}$$_{1}$$\mathit{T}$ follows a Curie-Weiss like the bulk
susceptibility. For Dirac quasiparticles (linear dispersion) in two-dimensions, the NMR spin-lattice relaxation rate (which depends on the square of the density-of-states) should show a power-law with $T$ (1/T$_{1}\propto T^{3}$). However, our measurements are only down to about 2 K. Perhaps, lower temperatures are required to observe this contribution. Another point is that NMR measurements are in an applied magnetic field which is also expected to affect the quasiparticle spectrum.

\section{$\mu$SR results}
The width of the Gaussian in the Kubo-Toyabe function from a fit of
the ZF data at 4.5 K is about 0.12 MHz. This amounts to a field of
about 1.4 Oe at the muon site which is a typical value for the field
from nuclear moments (Cu in this case). Next, data were taken in various
longitudinal fields to check how the muons decouple from the internal
fields. 

\begin{figure}[h]

\begin{centering}
\includegraphics[scale=0.35]{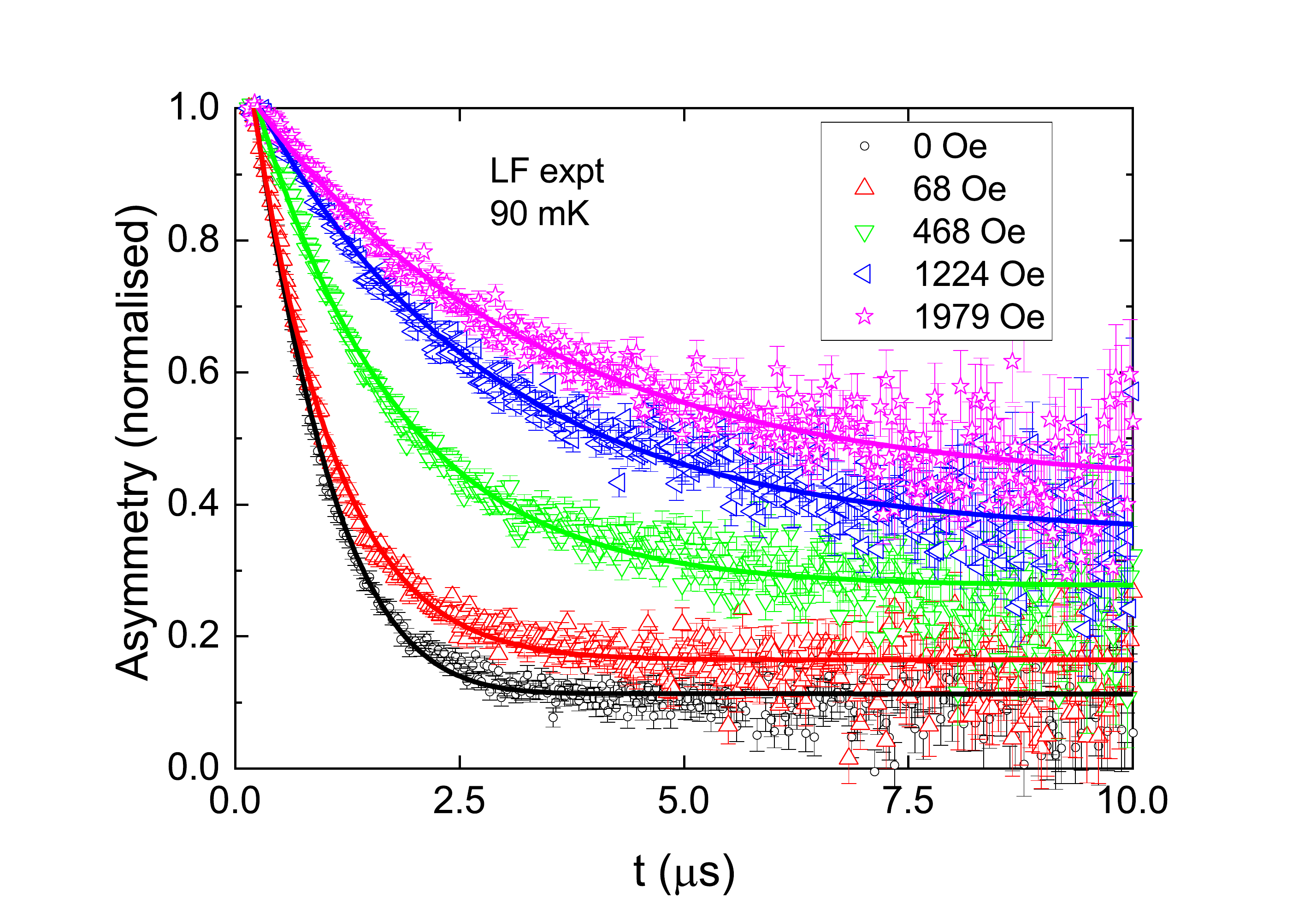}\caption{\label{fig:32}The variation of the muon asymmetry with time is shown
in some selected longitudinal fields at about 90 mK for Sr$_{3}$CuSb$_{2}$O$_{9}$.  The solid lines are fits as explained in the text.}
\par\end{centering}
\end{figure}

\begin{figure}[h]

\begin{centering}
\includegraphics[scale=0.35]{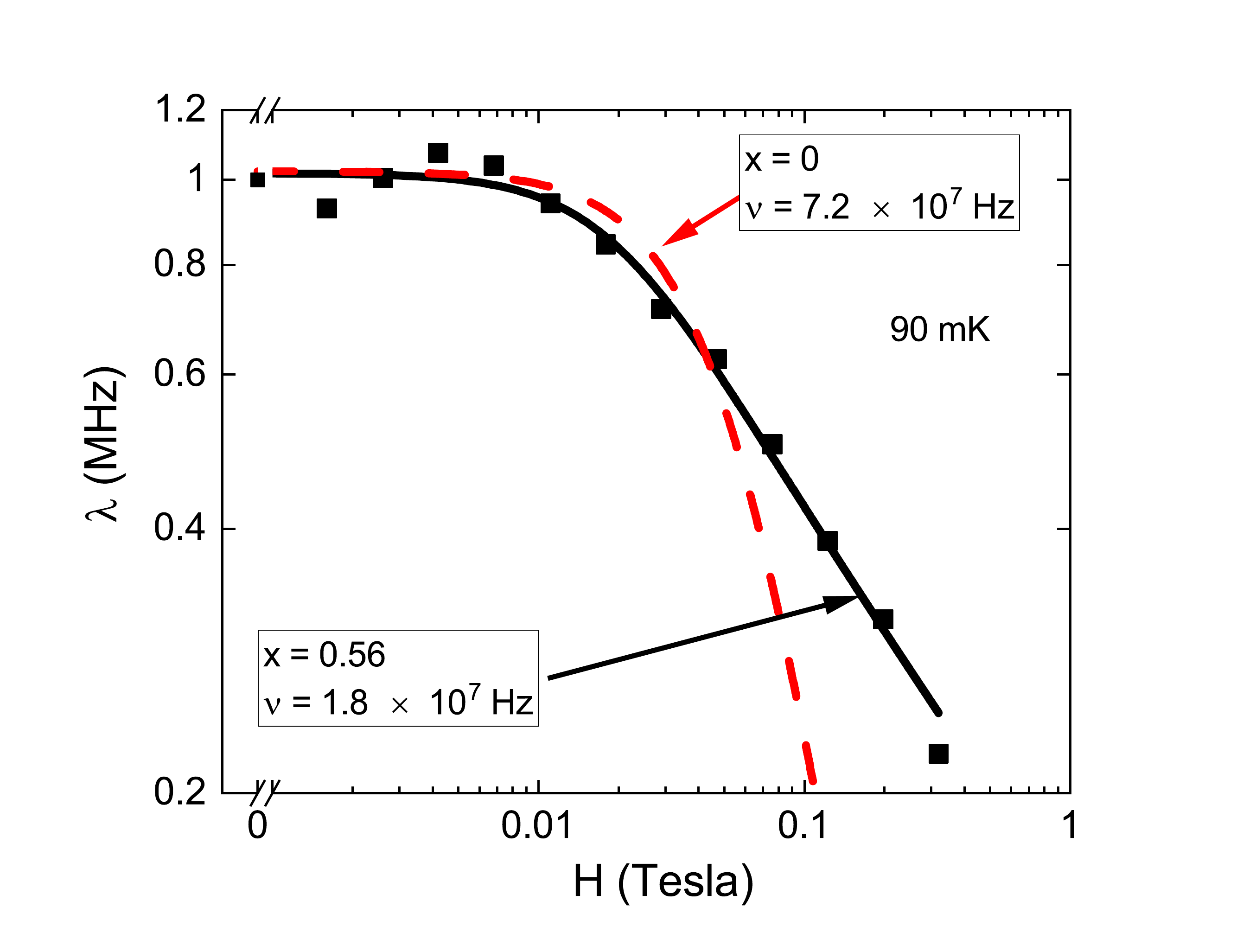}\caption{\label{fig:33}The variation of the muon relaxation rate at 90 mK
is shown as a function of the longitudinal magnetic field for Sr$_{3}$CuSb$_{2}$O$_{9}$.
The black curve is a fit to Equation \ref{eq:3} with x as a fitting parameter and the red dashed curve
is a fit to the same equation but with x = 0.}
\par\end{centering}
\end{figure}

From the data in Fig. \ref{fig:32}, it can be seen that even in a
large field of 1979 Oe, the muons are still not decoupled from the
internal fields. This indicates that the moments remain dynamic down
to the lowest temperatures studied. These are typical signatures seen
in quantum spin liquid materials. The LF data were also fit to the
product of the KT function with an exponential in addition to a constant
background. The muon depolarisation rate thus obtained is plotted
as a function of the magnetic field in Fig. \ref{fig:33}. 

At very high fields one expects muons to totally decouple from the internal fields resulting in no relaxation. But as seen in Fig. \ref{fig:33}, a field of 1979 Oe is not at all sufficient to make $\lambda$ near negligible. Following the analysis of the field dependence of $\lambda$ as in Ref. \cite{Li2016}, we fit the data to the following equation: 

\begin{equation}
\lambda (H) = 2\Delta ^2 \tau ^x  \intop _{0}^{\infty}t^{-x}exp(-\nu t)\cos (\gamma_{\mu}Ht) dt 	\label{eq:3}
\end{equation}
where $\nu$ is the fluctuation frequency of  local moments and $\Delta$
 is the distribution width of the local magnetic fields. The muon gyromagnetic ratio is $\gamma_{\mu}$ 
= 2$\pi$ x 135.5342 MHz/T.   A fit with $x = 0$ (red dashed curve in  Fig. \ref{fig:33}) which implies an exponential auto-correlation function $S(t) \sim exp (-\nu t)$ does not fit the data well and rather $S(t) \sim (\tau /t)^x exp (-\nu t)$ is needed to fit the data. The black solid curve is a
fit to Eq. \ref{eq:3} and gives $x = 0.56$ and $\nu = 1.8 \times 10^7 Hz $. $\tau$ is the early time cut-off and is fixed to 1 ps. This result implies the presence of long-time spin correlations but without any static order; a hallmark of spin liquids. While the numerical values above are similar to those found in YbMgGaO$_4$ \cite{Li2016}, we note that muon relaxation in SCSO is well fit to a single exponential (implying uniform relaxation for all muons) in contrast to YbMgGaO$_4$ where a stretched exponential behavior (implying a distribution of relaxation rates perhaps due to a distribution of environments) was found. 

\section{Details of Dirac Quantum Spin Liquid Ansatz}
For the Hamiltonian 
$\mathcal{H} = \sum_{\langle i,j \rangle \in \triangle} J \mathbf{S}_i \cdot \mathbf{S}_j
+ \ldots$
on the triangular lattice, where $\ldots$ stands for further neighbor interactions
providing frustration,
the following effective mean-field Hamiltonian is postulated as an anstaz 
\begin{align}
\mathcal{H} & = \frac{J_{\text{eff}}}{2} \sum_{\langle i,j \rangle \in \triangle \text{ lattice}, \sigma} 
\chi_{ij} c^\dagger_{i,\sigma} c_{j,\sigma} + \text{ h.c.} \nonumber \\
& = \frac{J_{\text{eff}}}{2} \sum_{\mathbf{k} \in \text{ B. Z.},\sigma} 
C^\dagger_{\mathbf{k},\sigma} \cdot H_{\mathbf{k}} \cdot C_{\mathbf{k},\sigma}
\end{align}
where
$C_{\mathbf{k},\sigma} = (c_{\mathbf{k},1,\sigma} \:\:\: c_{\mathbf{k},2,\sigma})^T$,
$c_{\mathbf{k},\mu,\sigma} = \frac{1}{N_{\text{UCs}}}\sum_{\mathbf{r}_\mu \in \square \text{ lattice}}
c_{\mathbf{r}_\mu,\sigma} e^{- i \mathbf{k} \cdot \mathbf{r}_\mu}$.
Here, the mean-field decoupling procedure for fermionic spin liquids (see Chapter 9 of Ref. \cite{Wen2007})
has been applied, and these fermionic degrees of freedom are termed as spinons.
$J_{\text{eff}}$  is an effective coupling parameter at the mean-field level. It is
expected to be of the order of $J_1$.
The ``$\square$" lattice in
the Fourier sum corresponds to the doubled unit cell that arises in the DSL ansatz
(See Fig. 3a of Ref. \onlinecite{Iqbal2016}). This gives rise to the $\mu$ index,
and the Fourier sum for $c_{\mathbf{k},\mu,\sigma}$ runs over only $\mu$ type of sites.
$\sigma$ is the $S=\frac{1}{2}$ index.
For the Dirac QSL ansatz defined via $\chi_{ij}$,
we follow the gauge choice of Ref. \onlinecite{Iqbal2016}.
The gauge-invariant fluxes still respect
the symmetry of the $\triangle$ lattice, such that
$\pi$-flux is present in the ``up" triangles, and
$0$-flux is present in the ``down" triangles.
The resulting lattice thus
is still chosen to have the same primitive lattice vectors as the underlying $\triangle$ lattice.
Going to Fourier space we arrive
at the matrix elements $h_{\mu \nu}(\mathbf{k})$ of the $H_{\mathbf{k}}$ matrix:
\begin{align}
h_{11}(\mathbf{k}) & = 2 \cos (\mathbf{k} \cdot \mathbf{a}_1) \nonumber \\
h_{22}(\mathbf{k}) & = - 2 \cos (\mathbf{k} \cdot \mathbf{a}_1) \nonumber \\
h_{12}(\mathbf{k}) & = 2 \cos (\mathbf{k} \cdot \mathbf{a}_2) - 
2 i \sin(\mathbf{k} \cdot (\mathbf{a}_1+\mathbf{a}_2)) \nonumber \\
h_{21}(\mathbf{k}) &= h_{12}(\mathbf{k})^*
\end{align}
or compactly, $H_{\mathbf{k}} = 2 \cos (\mathbf{k} \cdot \mathbf{a}_1) \tau_z
+ 2 \cos (\mathbf{k} \cdot \mathbf{a}_2) \tau_x 
+ 2 \sin(\mathbf{k} \cdot (\mathbf{a}_1+\mathbf{a}_2)) \tau_y$ where 
the $\tau$ Pauli matrices operate on the $\mu$ indices, $\mathbf{a}_1=(1,0)$,
$\mathbf{a}_2=(-\frac{1}{2},\frac{\sqrt{3}}{2})$ in units of the lattice constant
(which we take to be dimensionless unity).
Diagonlizing $H_{\mathbf{k}}$, we arrive at a single Dirac cone in the Brillouin Zone
with a choice of primitive reciprocal lattice vectors as
$\mathbf{b}_1 = 2 \pi (1, \frac{1}{\sqrt{3}})$, 
$\mathbf{b}_2 = 2 \pi (0, \frac{2}{\sqrt{3}})$ in units of inverse lattice constant. 
The eigenvalues of $H_{\mathbf{k}}$ are
$\pm \sqrt{2} \left( \frac{J_{\text{eff}}}{2} \right) \left[3 + \cos(2 k_x) + 2 \sin(k_x) \sin(\sqrt{3} k_y) \right]^{1/2}$.
Expanding near the Dirac cone at $\mathbf{k}=(\frac{\pi}{2}, \frac{\sqrt{3}\pi}{2})$, 
we arrive the effective low energy spectrum
$\pm \epsilon_{\mathbf{k}} \equiv \pm \epsilon_{k} = \pm \sqrt{6} \left( \frac{J_{\text{eff}}}{2} \right) k$.
Note $k$ is dimensionless since we 
take the lattice constant to be dimensionless unity.
We have to stay at half-filling 
(on average at mean-field level). This is ensured by the zero chemical potential
for this particle-hole symmetry spectrum.

With the low energy spectrum in hand, we can work out the specific heat contribution
from the Dirac QSL:
\onecolumngrid
\begin{align}
E(T) - E(T=0) & 
= 2 \left[ \int^{\Lambda_k}_0 \frac{d^2 \mathbf{k}}{\mathcal{A}_{\text{UC}}
	\mathcal{A}_{\text{BZ}}} f(\epsilon_{\mathbf{k}}) (+\epsilon_{\mathbf{k}})  
+ \int^{\Lambda_k}_0 \frac{d^2 \mathbf{k}}{\mathcal{A}_{\text{UC}}\mathcal{A}_{\text{BZ}}}
(f(-\epsilon_{\mathbf{k}})-1) (-\epsilon_{\mathbf{k}})
\right] \nonumber \\
& = \frac{2}{\pi} \int^{\Lambda_k}_0 k dk \; \frac{\epsilon_k}{e^{\beta \epsilon_k}+1}
= \frac{1}{3\pi \left( \frac{J_{\text{eff}}}{2} \right)^2} \int^{\sqrt{6} \left( \frac{J_{\text{eff}}}{2} \right) \Lambda_k }_0 d\epsilon \; \frac{\epsilon^2}{e^{\beta \epsilon}+1}
\nonumber \\
& = \frac{1}{3\pi \left(\frac{J_{\text{eff}}}{2}\right)^2 \beta^3} 
\int^{\sqrt{6} \left( \frac{J_{\text{eff}}}{2} \right) \Lambda_k \beta}_0 
dy \; \frac{y^2}{e^{y}+1} = 
\frac{1}{3\pi \left(\frac{J_{\text{eff}}}{2}\right)^2 \beta^3} \int^{\infty}_0 
dy \; \frac{y^2}{e^{y}+1} \: \text{ as }\beta \rightarrow \infty \nonumber \\
& \approx 0.191 \frac{(k_B T)^3}{\left(\frac{J_{\text{eff}}}{2}\right)^2}  
\label{eq:algebra} \\
\implies \frac{C(T)}{k_B} & \approx 2.292 \frac{(k_B T)^2}{J_{\text{eff}}^2} 
\end{align}
\twocolumngrid
The factor of 2 outside the big square brackets in the first line above
is due to summing over the spin quantum number $\sum_\sigma$.
We may include an external applied field to the above analysis via a
Zeeman coupling to the spins, i.e. $\mathcal{H} = 
J \sum_{\langle i,j \rangle \in \triangle} \mathbf{S}_i \cdot \mathbf{S}_j
+ g \mu_B \sum_i \mathbf{H} \cdot \mathbf{S}_i$. Again making the Dirac QSL ansatz
and taking the spin quantization axis along the external magnetic field 
$\mathbf{H}$,
we arrive at
\onecolumngrid
\begin{align}
\mathcal{H} & = \frac{J_{\text{eff}}}{2} \sum_{\langle i,j \rangle \in \triangle \text{ lattice}, \sigma} 
\chi_{ij} c^\dagger_{i,\sigma} c_{j,\sigma} + \text{ h.c.} 
+ H' \sum_i \left( c^\dagger_{i,\uparrow} c_{j,\uparrow}
- c^\dagger_{i,\downarrow} c_{j,\downarrow} \right) \nonumber \\
& = \frac{J_{\text{eff}}}{2} \sum_{\mathbf{k} \in \text{ B. Z.},\sigma} 
C^\dagger_{\mathbf{k},\sigma} \cdot H_{\mathbf{k},\sigma} \cdot C_{\mathbf{k},\sigma}
\end{align}
\twocolumngrid
where $H'$ is a lumped parameter with same dimensions
as $J_{\text{eff}}$ ($H' = g \mu_B |\mathbf{H}|/2$).
Going to Fourier space as before, we arrive
at the matrix elements $h_{\mu \nu, \sigma}(\mathbf{k})$ 
of the $H_{\mathbf{k},\sigma}$ matrix:
\begin{align}
h_{11,\sigma}(\mathbf{k}) & = 2 \cos (\mathbf{k} \cdot \mathbf{a}_1) 
+ (-1)^\sigma \frac{H'}{\left( \frac{J_{\text{eff}}}{2} \right)} \nonumber \\
h_{22,\sigma}(\mathbf{k}) & = - 2 \cos (\mathbf{k} \cdot \mathbf{a}_1) 
+ (-1)^\sigma \frac{H'}{\left( \frac{J_{\text{eff}}}{2} \right)} \nonumber \\
h_{12,\sigma}(\mathbf{k}) & = 2 \cos (\mathbf{k} \cdot \mathbf{a}_2) - 
2 i \sin(\mathbf{k} \cdot (\mathbf{a}_1+\mathbf{a}_2)) \nonumber \\
h_{21,\sigma}(\mathbf{k}) &= h_{12}(\mathbf{k})^*
\end{align}
or compactly, $H_{\mathbf{k},\sigma} = 2 \cos (\mathbf{k} \cdot \mathbf{a}_1) \tau_z
+ 2 \cos (\mathbf{k} \cdot \mathbf{a}_2) \tau_x 
+ 2 \sin(\mathbf{k} \cdot (\mathbf{a}_1+\mathbf{a}_2)) \tau_y
+ (-1)^\sigma \frac{H'}{\left( \frac{J_{\text{eff}}}{2} \right)} \tau_0$. It is understood that $(-1)^\uparrow=1$,
$(-1)^\downarrow=-1$ and $\tau_0$ is the identity matrix in the $\mu$ index.
Going through the same steps as before, we will now arrive at
the effective low energy spectrum
$\pm \epsilon_{\mathbf{k},\sigma} 
\equiv \pm \epsilon_{k,\sigma} = \pm \epsilon_k + (-1)^\sigma H'$,
i.e. the applied field acts like a chemical potential with opposite sign
for $\sigma=\uparrow$ and $\sigma=\downarrow$.
With the low energy spectrum in hand, we can work out the specific heat 
contributions due to imbalanced occupations of the
two spin species.\cite{chem-pot}


The energy contributions are shown below:
\onecolumngrid
\begin{align}
E_\uparrow(T) - E_\uparrow(T=0) & 
=  \Bigg[ \int^{\Lambda_k}_0 \frac{d^2 \mathbf{k}}{\mathcal{A}_{\text{UC}}
	\mathcal{A}_{\text{BZ}}} f(\epsilon_{\mathbf{k},\uparrow}) 
(+\epsilon_{\mathbf{k},\uparrow})  
+ \int^{k|_{\epsilon_k = H'}}_0 \frac{d^2 \mathbf{k}}{\mathcal{A}_{\text{UC}}
	\mathcal{A}_{\text{BZ}}} f(-\epsilon_{\mathbf{k},\uparrow}) 
(-\epsilon_{\mathbf{k},\uparrow}) 
\nonumber \\
& \; \; \; \; + \int^{\Lambda_k}_{k|_{\epsilon_k = H'}} \frac{d^2 \mathbf{k}}{\mathcal{A}_{\text{UC}}\mathcal{A}_{\text{BZ}}}
(f(-\epsilon_{\mathbf{k},\uparrow})-1) (-\epsilon_{\mathbf{k},\uparrow})
\Bigg] \\
E_\downarrow(T) - E_\downarrow(T=0) & 
=  \Bigg[ \int^{\Lambda_k}_{k|_{\epsilon_k = H'}} \frac{d^2 \mathbf{k}}{\mathcal{A}_{\text{UC}}
	\mathcal{A}_{\text{BZ}}} f(\epsilon_{\mathbf{k},\downarrow}) 
(+\epsilon_{\mathbf{k},\downarrow})  
+ \int^{k|_{\epsilon_k = H'}}_0 \frac{d^2 \mathbf{k}}{\mathcal{A}_{\text{UC}}
	\mathcal{A}_{\text{BZ}}} (f(\epsilon_{\mathbf{k},\downarrow})-1) 
(\epsilon_{\mathbf{k},\downarrow}) 
\nonumber \\
& \; \; \; \; + \int^{\Lambda_k}_{0} \frac{d^2 \mathbf{k}}{\mathcal{A}_{\text{UC}}\mathcal{A}_{\text{BZ}}}
(f(-\epsilon_{\mathbf{k},\downarrow})-1) (-\epsilon_{\mathbf{k},\downarrow})
\Bigg] \nonumber 
\end{align}
\twocolumngrid
For the above, we again follow similar steps as in Eq. \ref{eq:algebra}
to arrive at the sum of the two contributions:
\onecolumngrid 
\begin{equation}
E(T,H) - E(T=0,H) \approx 0.191 \frac{(k_B T)^3}{\left( \frac{J_{\text{eff}}}{2} \right)^2}
+ \frac{(k_B T)^3}{3 \pi \left( \frac{J_{\text{eff}}}{2} \right)^2} \int^{\beta H'}_0 dy \frac{(\beta H' - y)y}
{e^y+1}
\end{equation}
\twocolumngrid
In the above, we have assumed $\beta J_{\text{eff}} \rightarrow \infty$ similar to the
zero field calculation in Eq. \ref{eq:algebra}, while no such assumption is put on $\beta H'$.
When $\beta H' \rightarrow 0$, we recover the $T^2$ behaviour in
the specific heat as expected. When $\beta H' \rightarrow \infty$,
we will get in the limiting case
\onecolumngrid
\begin{align}
E(T,H) - E(T=0,H) & \approx 0.191 \frac{(k_B T)^3}{\left( \frac{J_{\text{eff}}}{2} \right)^2}
+ \frac{(k_B T)^2 H'}{3 \pi \left( \frac{J_{\text{eff}}}{2} \right)^2} \int^{\infty}_0 dy \frac{y}{e^y+1}
\nonumber \\
& \approx 0.191 \frac{(k_B T)^3}{\left( \frac{J_{\text{eff}}}{2} \right)^2}
+ 0.087 \frac{(k_B T)^2 H'}{\left( \frac{J_{\text{eff}}}{2} \right)^2} \nonumber \\
\implies \frac{C(T,H)}{k_B} & \approx 2.292 \frac{(k_B T)^2}{J_{\text{eff}}^2}
+ 0.696 \frac{(k_B T) H'}{J_{\text{eff}}^2}
\label{eq:sphtwH}
\end{align}
\twocolumngrid
We have neglected the $-\frac{y \left(\frac{y}{\beta H'}\right)}{e^y + 1}$
piece of the integrand in 
the second term's integral above, since it goes as $\frac{1}{\beta H'} \rightarrow 0$
for small $y$, and $\beta H' e^{-\beta H'} \rightarrow 0$ for large $y$.
Thus in the $\beta H' \rightarrow \infty$ limit, we now have a $T$-linear behaviour
in the specific heat. This is also in order, because once the temperature
becomes much smaller than the energy scale associated with the magnetic field, then the spinon
occupations effectively correspond to a situation with a fermi surface,
where a $T$-linear behaviour in the specific
heat is expected. So, in the presence of the magnetic field, the specific
heat interpolates between a $T$-linear behaviour as $\beta H' \rightarrow \infty$
to a $T^2$ behaviour as $\beta H' \rightarrow 0$. The coefficients
in Eq. \ref{eq:sphtwH} are not considered 
sacrosanct when used as a fitting form for experimental data, since they
were arrived at using several assumptions. But the $T$-dependence interpolation
in presence of an applied field is taken seriously when using Dirac QSL ansatz for
phenomenology.

\section{Electronic Structure calculation}
The electronic structure calculations were done for an ordered crystal structure, such that within the unit cell of Sr$_3$CuSb$_2$O$_9$, Cu atoms are connected through
two consecutive Sb atoms forming a Cu-Cu triangular network within (111) plane. To construct such a crystal structure we have 
considered an ordered 3$\times$3$\times$3 super-cell (containing 270 atoms) of the two formula unit primitive unit cell of the parent
compound Sr(Cu,Sb)O$_3$. The triangular Cu layers on the (111) plane are connected via Cu-O-Sb-O-Sb-O-Cu path. The distances between 
the six nearest neighbor Cu atoms in the triangular network are nearly identical. The local CuO$_6$ octahedral environment breaks the 
degeneracy of Cu-d orbitals into triply degenerate t$_{2g}$ and doubly degenerate e$_g$ orbitals. Our theoretically calculated non spin-polarized  total and Cu-d partial density of 
states (DOS) plot is shown below.
\begin{figure}[h]
	\begin{centering}
		\includegraphics[scale=0.5]{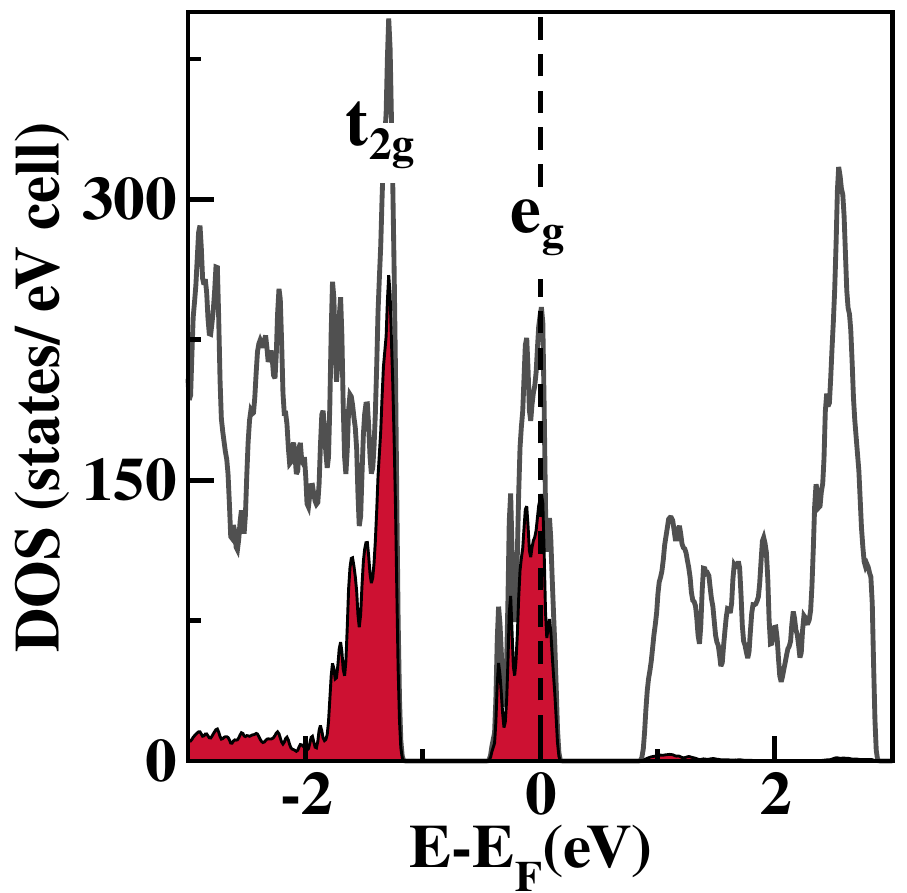}\caption{\label{fig:35} Non-spin polarized total (grey) and Cu-d partial (red) DOS for SCSO.}
		\par\end{centering}
\end{figure}
\bibliographystyle{apsrev4-1}
\bibliography{SCSO}